\documentclass{aastex62}
\pdfoutput=1 
\usepackage{subfigure}

\graphicspath{{./}}

\shorttitle{Comparison of the physical properties of the L4 and L5 Trojans from ATLAS}
\shortauthors{McNeill et al.}

\begin{document}

\title{Comparison of the physical properties of the L4 and L5 Trojan asteroids from ATLAS data}

\correspondingauthor{Andrew McNeill}
\email{andrew.mcneill@nau.edu}

\author{A. McNeill}
\affiliation{Department of Astronomy and Planetary Science, Northern Arizona University, Flagstaff, AZ 86011, USA}

\author{N. Erasmus}
\affil{South African Astronomical Observatory, Cape Town, 7925, South Africa.}

\author{D.E. Trilling}
\affiliation{Department of Astronomy and Planetary Science, Northern Arizona University, Flagstaff, AZ 86011, USA}
\affil{South African Astronomical Observatory, Cape Town, 7925, South Africa.}

\author{J.P. Emery}
\affiliation{Department of Astronomy and Planetary Science, Northern Arizona University, Flagstaff, AZ 86011, USA}

\author{J. L. Tonry}
\affil{Institute for Astronomy, University of Hawaii, Honolulu, HI 9682, USA.}

\author{L. Denneau}
\affil{Institute for Astronomy, University of Hawaii, Honolulu, HI 9682, USA.}

\author{H. Flewelling}
\affil{Institute for Astronomy, University of Hawaii, Honolulu, HI 9682, USA.}

\author{A. Heinze}
\affil{Institute for Astronomy, University of Hawaii, Honolulu, HI 9682, USA.}

\author{B. Stalder}
\affil{Vera C. Rubin Observatory Project Office, 950 N. Cherry Ave, Tucson, AZ, USA}

\author{H.J. Weiland}
\affil{Institute for Astronomy, University of Hawaii, Honolulu, HI 9682, USA.}

\begin{abstract}
Jupiter has nearly 8000~known co-orbital asteroids orbiting in the L4 and L5 Lagrange points called Jupiter Trojan asteroids.
Aside from the greater number density of the L4 cloud the two clouds are in many ways considered to be identical. Using sparse photometric data taken by the Asteroid Terrestrial-impact Last Alert System (ATLAS) for 863 L4 Trojans and 380 L5 Trojans we
derive the shape distribution for each of the clouds
and find that, on average,
the L4 asteroids are more elongated than the L5 asteroids.
This shape difference is most likely due to the 
greater collision rate in the L4 cloud that results from its larger population.
We additionally present the phase functions and $c-o$ colours of 266~objects.

\end{abstract}


\section{Introduction} \label{sec:intro}

Jupiter Trojans are minor planets 
that orbit 60~degrees ahead of (L4) and behind (L5) Jupiter in the 1:1~resonant
Lagrange points.
As of January, 2020, the number of known Trojans listed by the Minor Planet Center is 7673, with \cite{nakamura2008} predicting a total of $10^{5}$ Trojans with diameter ($D$) greater than 2~km across the two clouds. Of the Trojans listed in the MPC, 4952 orbit as part of the L4 cloud and 2721 orbit in the L5.

Our understanding of the Trojan population will be greatly enhanced by the forthcoming Lucy mission, which will explore targets in both clouds starting in 2025. The Nice model (\citealt{gomes2005}) predicts that the current Trojan population may have formed much further from the Sun than their current location and may consist of material scattered from the outer Solar System that is subsequently captured by Jupiter. This formation mechanism does not reproduce the difference in number between the two Trojan clouds: The L4 cloud contains a greater number of objects with $D > 10$ km than the L5 cloud by a factor of approximately 1.4 (\citealt{grav2011}). \cite{nesvorny2013} consider a capture mechanism involving excitation 
of Jupiter's orbit 
in the early evolution of the Solar System
due to the presence of a fifth giant planet. During the orbital instability resulting from encounters between Jupiter and this fifth planet, Jupiter's position and by extension its Lagrangian points move, resulting in the loss of primordial Trojans previously occupying these stable regions. The L4 and L5 clouds are then repopulated with material captured in the post-migration orbit of Jupiter. This material is proposed to be similar in origin to current outer Solar System (e.g., Kuiper Belt Object) populations, although modelling by \cite{nesvorny2013} predicts that most of the material was captured close to Jupiter's current orbit. The L4/L5 number asymmetry can be explained in this case if a giant planet passes through the L5 cloud in its motion leading to this cloud being preferentially depleted (\citealt{nesvorny2013}).

In addition to a difference in the number of objects in each cloud 
the remaining properties of Trojan asteroids (for instance, the observed colour/spectral dichotomy between less red and more red populations)
also show some difference between the clouds. \cite{szabo2007} observe a colour-dependence on inclination in both clouds independent of size, and match a power law to both distributions. The difference in colour distribution between the clouds was explained as potentially due to the different number densities and could be solved by normalising each cloud differently. \cite{roig2008} observe a clearly different distribution of spectral slopes between the two clouds suggesting that this is due to the presence of collisional families in the clouds, as considering only background objects yields identical distributions in both L4 and L5. \cite{emery2011} discovered a bimodality in the spectral slopes of Jupiter Trojans between 'red' and 'less-red' objects, which correlates with the colour bimodality observed at visible wavelengths. This effect is seen equally in both L4 and L5 clouds.


In this paper we present a brief description of ATLAS and the data
used in Section 2 and an overview of the shape distribution model used
in Section 3. We present and discuss the colour measurements, phase functions and rotation periods derived for Trojans in ATLAS in Sections 4 and 5
respectively.


\section{Asteroid Terrestrial-impact Last Alert System (ATLAS)} \label{sec:ATLAS}

The photometry data used in this study originate from survey observations performed between 2015 and 2018 by the Asteroid Terrestrial-impact Last Alert System (ATLAS)\footnote[1]{\url{http://atlas.fallingstar.com}}. Currently consisting of two units both located in Hawai`i, ATLAS is designed to achieve a high survey speed per unit cost \citep{Tonry2018}. Its main purpose is to discover asteroids with imminent impacts with Earth that are either regionally or globally threatening in nature. To fulfill this, the two current ATLAS units scan the complete visible northern sky every night enabling it to make numerous discoveries in multiple astronomical disciplines, such as supernovae candidates discovery \citep{Prentice2018}, gamma ray burst phenomena \citep{Stalder2017}, variable stars \citep{Heinze_2018}, and asteroid discovery \citep{Tonry_2018b}. All detected  asteroid astrometry and photometry are posted to the Minor Planet Center, while the supernova candidates are publicly reported to the International Astronomical Union Transient Name Server. The 5-sigma limiting magnitude (AB) per 30 sec exposure is 19.7 for both ATLAS filters.

\begin{figure}
\begin{center}
\includegraphics[width=0.5\textwidth]{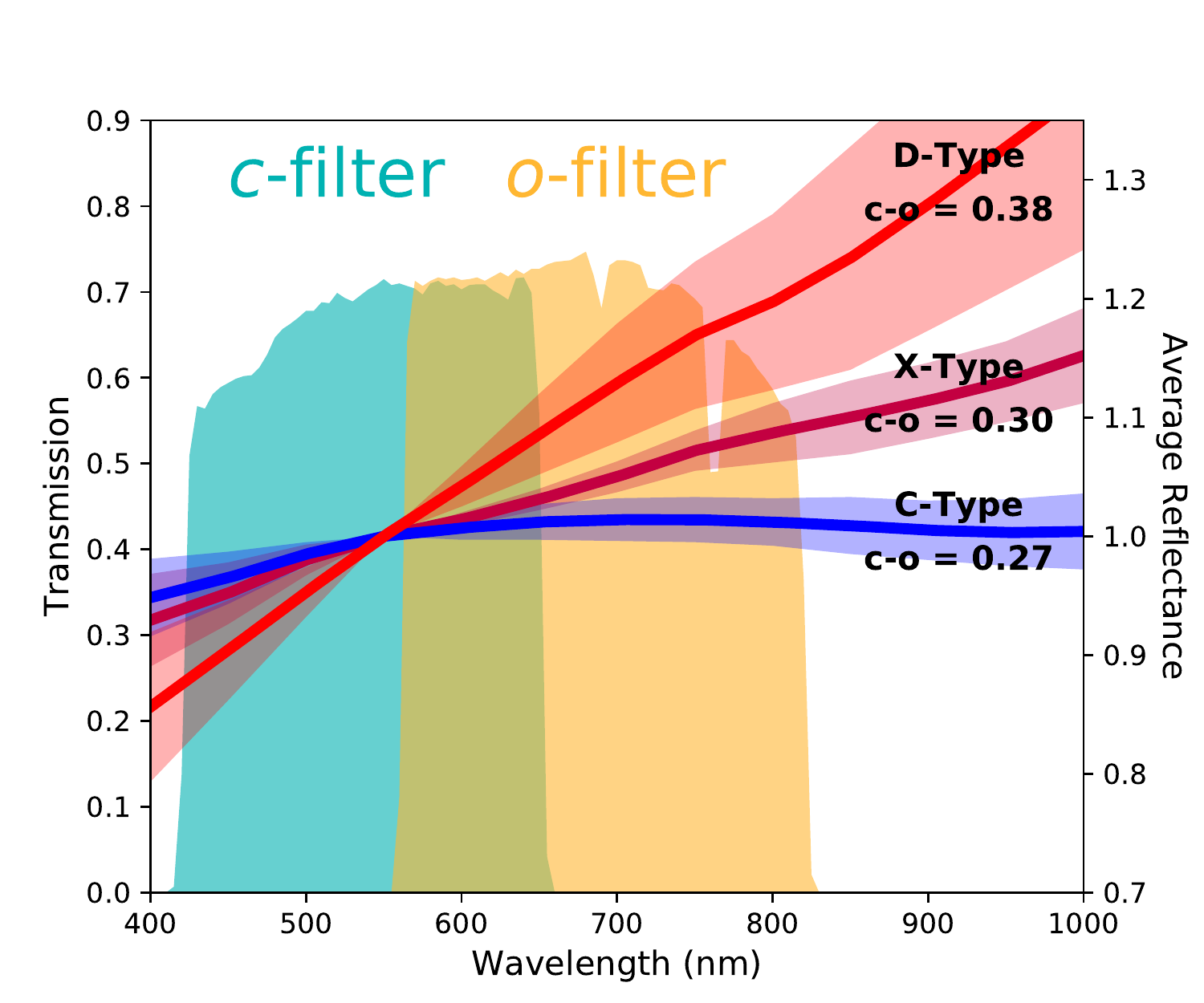}
\caption{The transmission curves of the \textit{c}- and \textit{o}-filters of ATLAS \citep{Tonry2018} are plotted together with the averaged visible wavelength reflectance spectra, normalized at 550\,nm, of the featureless Bus-DeMeo C-, X- and D-type  taxonomies \citep{DeMeo2009}. The upper and lower bounds of mean spectra provided by \cite{DeMeo2009} are also indicated with shading. The \textit{c-o} colour of each taxonomic complex is calculated by convolving the
ATLAS filter responses with the mean Bus-DeMeo spectra. The expected colour values are also shown.}
\label{filter_spectra}
\end{center}
\end{figure}

The two ATLAS units are 0.5\,m telescopes each covering 30 deg$^2$ field-of-view in a single exposure. The main survey mode mostly utilizes two custom filters, a ``cyan'' or \textit{c}-filter with a bandpass between 420--650\,nm and an ``orange'' or \textit{o}-filter with a bandpass between 560--820\,nm (as shown in Figure~\ref{filter_spectra}). For further details on ATLAS, ATLAS photometry, and the ATLAS All-Sky Stellar Reference Catalog see \cite{Tonry2018,Heinze_2018} and \cite{Tonry_2018b} . Although this AB photometric system uses only two, relatively wide filters the $c-o$ colour obtained from ATLAS detections can be a good initial diagnostic to distinguish among asteroid taxonomic types. Further detail on this methodology can be read in \cite{erasmus2020}.

For this project, 
photometry data of 863 L4 Trojans and 380 L5 Trojans could be extracted from the ATLAS data set as of this writing. We typically have a median of 55 and 74 unique observations for each detected L4 and L5 Trojan respectively, with roughly 30\% in $c$ and 70\% in $o$ band for both clouds. For this study we limit the data set to objects that had at least 100 and 20 observations in the \textit{o}- and\textit{c}-filter respectively. We also only considered objects that had at least one observation at a phase angle (Sun-Observer-Target angle), $\alpha$, of 5 degrees or lower and phase angle coverage of at at least 6 degrees to ensure reliable phase curve fits. These criteria distilled the ATLAS datset to 209 L4 Trojans and 133 L5 Trojans for which we had a median of $\sim$300 observations, a median for the minimum phase angle of $1.3$$^{\circ}$, and a median phase angle range of $9.6$$^{\circ}$. For all analysis in this work, we first cast all observed magnitudes to a corresponding absolute H$_{c}$ and H$_{o}$ magnitude
by removing the distance and phase angle dependence for all objects and for both filters. Hereafter, all references to o- and c-filter data refers to the H$_{c}$ and H$_{o}$ magnitudes unless stated otherwise.


\section{Shape Distribution Model} 

The statistical shape model employed here generates a synthetic population of triaxial ellipsoids with assumed shapes and spin pole orientations based on input distributions. We generate synthetic observations of these objects at an observing cadence equivalent to that of the ATLAS survey.
We then compare the resulting set of individual observations of synthetic objects from different shape and spin-state distributions to the observed data using the two-sampled Kolmogorov-Smirnov test as well as Mann-Whitney and chi-squared fits for confirmation. Only relative changes in brightness due to geometry are considered; as such we do not need to account for heliocentric distance or surface characteristics. We considered applying the method of \citealt{mcneill2018} used for NEOs in Spitzer observations, however, to compare sets of rotational amplitudes rather than individual detections requires observations made contiguously rather than sparsely over a long range of time.

Previous distribution models applied to main belt asteroids and Near-Earth Objects have assumed a 
uniform spin frequency distribution from 1.0--10.9 day\textsuperscript{-1} across all applicable size ranges, corresponding to rotational periods from the spin barrier at 2.2~h to 24~h (\citealt{mcneill2016}; \citealt{mcneill2018}).
This assumption is reasonable given the flat distribution of measured rotational frequencies at small asteroid sizes \citep{pravec2002}. For the Trojan populations, we also include a population of slow rotators. 







For generated synthetic detections a uniformly distributed uncertainty value is selected between -0.05 and 0.05 magnitudes, consistent with the uncertainty values in the selected ATLAS data set and applied to the value. 
We do not include any effect of limb scattering and/or darkening, which would only be significant at phase angles larger than the maximum of $10^{\circ}$ reached by Trojans.

As we are only concerned with the relative magnitude differences caused by differing shapes, the value of $a$ can be fixed and the values of $b$ and $c$ can be varied using various distributions e.g. Gaussians, Lorentzians, bimodal distributions. A range of synthetic populations are generated from different input shape and spin pole distributions and compared to the observational data using the two-sample Kolmogorov-Smirnov and Mann-Whitney tests. Identical distributions would produce a value of $p=1$ in these cases. The results of these statistical tests indicate how closely the synthetic population resembles the underlying population that ATLAS sampled.

In order to test the validity of the fits from our model, we applied it to a series of known distributions. We first generated a range of known distributions of the axis ratio $b/a$, truncated at $b/a=1$ since $b$ cannot be greater than $a$. We assume $a=1$ in all cases. From each input shape distribution, pseudo-ATLAS data was generated accounting for the cadence of ATLAS observations and the typical phase-angle distribution for Jupiter Trojans. The model was applied to these distributions and the returned best fit values compared with the known parameters. Using multiple two-sampled statistical tests we find that the derived result is correct in mean aspect ratio, and Gaussian parameters where a Gaussian distribution was assumed. The width of this Gaussian is a free parameter ranging from 0.1-0.4, small variations in this and the centre of the Gaussian distribution produce equally good fits. The mean elongation for these best fits remains consistent, however, and we favour it as a metric in this analysis.

\section{Results} \label{sec:results}

\subsection{Colours and Phase Curves}\label{sec:colours}

\begin{figure*}
\gridline{\fig{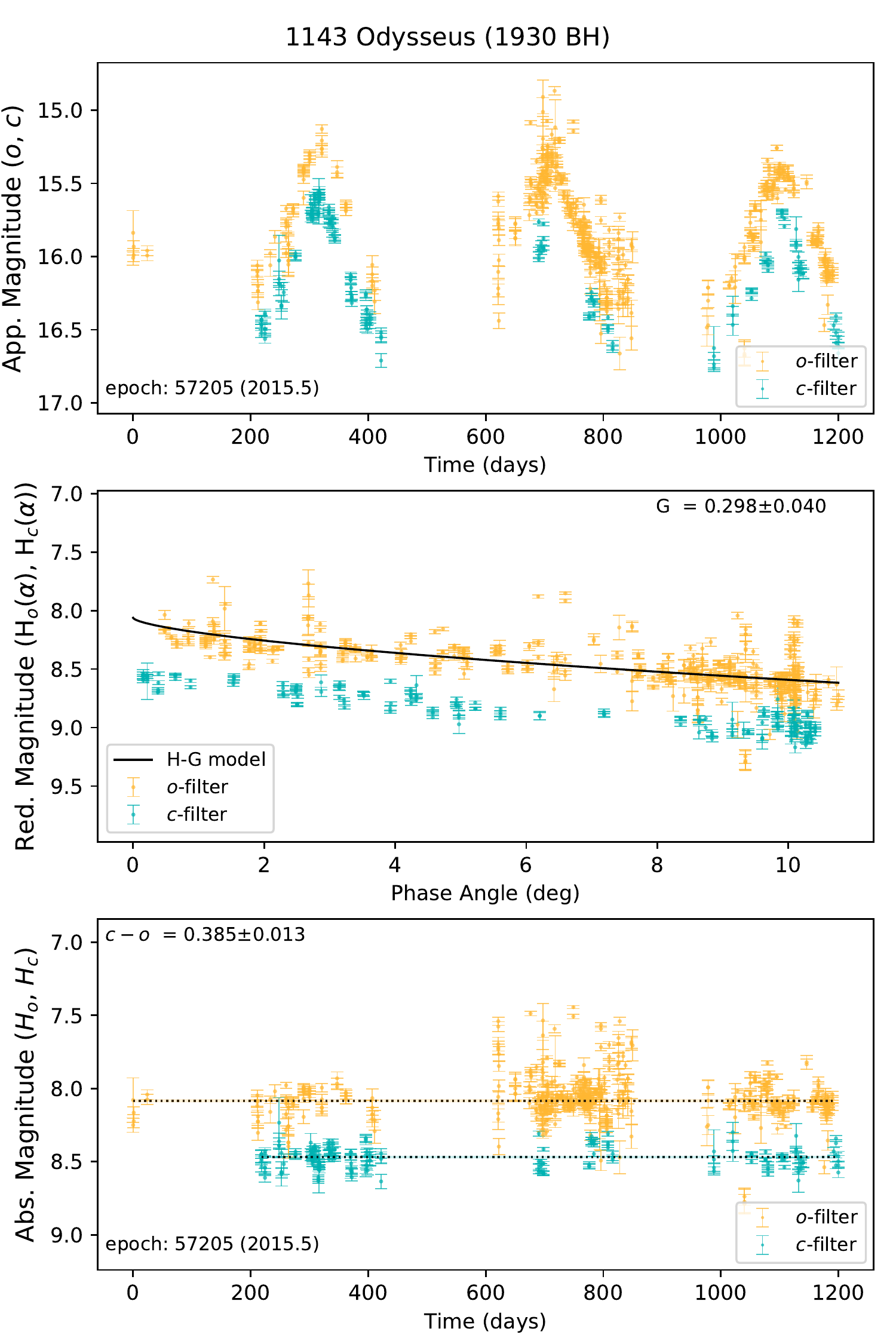}{0.45\textwidth}{(a)}\fig{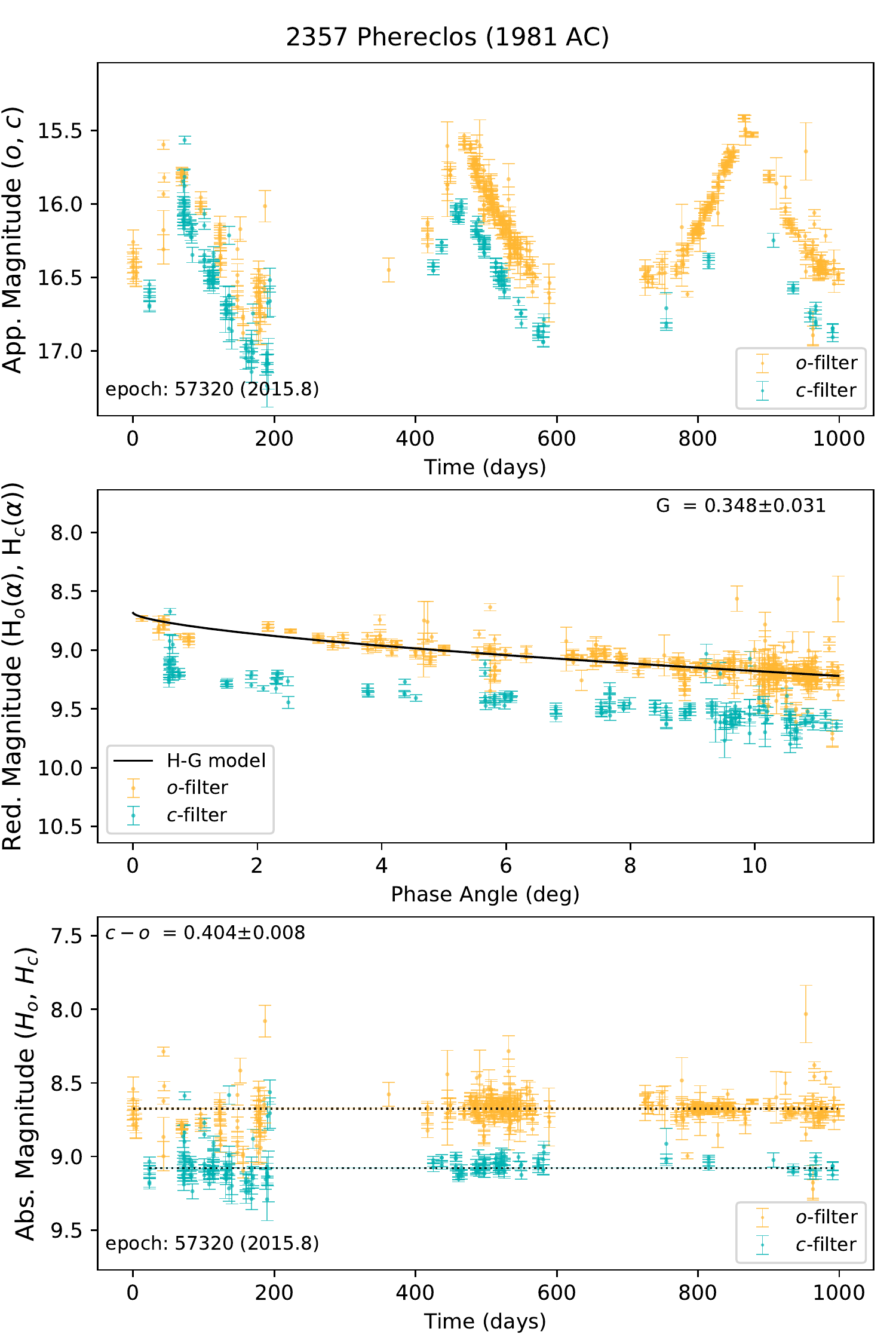}{0.45\textwidth}{(b)}}
\caption{Example ATLAS photometric data for (a) L4 Trojan 1143 Odysseus (1930 BH) and (b) L5 Trojan 2357 Phereclos (1981 AC). The first panel in each shows the observed magnitude spanning several oppositions. The second panel shows the reduced magnitude as a function of observed phase angle with the H-G phase curve model fit and the fitted phase parameter indicated in the top-right corner. The bottom panel for each object show absolute magnitudes H$_{c}$ and H$_{o}$ with the gap between the horizontal dotted lines the extracted \textit{c-o} color (see Section \ref{sec:colours}) with the value also indicated in the top-left corner of the panel.} 
\label{L4_L5_example_photometry}
\end{figure*}

To determine the colours of each ATLAS object we first cast all observed magnitudes of both the \textit{c}- and \textit{o}-filter data (see top panel of Figure \ref{L4_L5_example_photometry}) to reduced magnitude, $H(\alpha)$, by removing the influence of distance on the brightness of the body. This reduced magnitude depends only on the phase angle, $\alpha$, of the observations Using the formulation by \citet{Bowell1989} we fit the H-G model to the reduced magnitudes to extract a fitted phase curve parameter (see middle panel of Figure \ref{L4_L5_example_photometry} and values recorded in Table \ref{table_col_data}). The improved H-G1/G2 (cite Muinonen) better fits sparse data that spans both very small and large phase angles and therefore also better fits the the most dramatic part of opposition surge in brightness at phase angle equal to zero. However, for this study we have opted for the less-complex H-G model since we have many observations for each object but those observations do not span a very large phase angle range ($<10$ degrees, see Section \ref{sec:ATLAS}) We also do not have any observations for most objects at a zero degree phase angle. Our fitted H-G model is also used to cast all reduced magnitudes to absolute magnitudes H$_{c}$ and H$_{o}$ (see bottom panel of Figure \ref{L4_L5_example_photometry}) which are used for color determination and for the shape modeling analysis (see Section 4.3). 

A large contribution to the remaining scatter in the H$_{c}$ and H$_{o}$ magnitudes is the potential brightness variation due to the asteroid's rotation so the \textit{c-o} colour of each object is extracted by determining the median  H$_{c}$ and H$_{o}$ magnitudes of a randomly selected 50\% subset of the \textit{c}- and \textit{o} filter data, respectively, and repeating that process 10 times and calculating the the average. The final \textit{c-o} colour is defined as the difference between the average of the median magnitudes.  The uncertainty in the colour value incorporates the standard deviation of the median magnitudes in the previous step. The bottom plots of Figure \ref{L4_L5_example_photometry} show examples of the results of this procedure with the \textit{c-o} colour values and uncertainty displayed in top-left corner of each plot and recorded in Table \ref{table_col_data}.\\

\subsection{Rotation Periods}\label{sec:periods}
Rotation periods were extracted from the ATLAS data by generating Lomb-Scargle periodograms \citep{lomb1976,scargle1982} of each target's \textit{o}-filter photometric data (as there are more \textit{o} measurements than \textit{c} measurements). Targets that generated periodograms containing a peak with a false-alarm probability $\lesssim 10^{-10}$ were flagged to have a potentially extractable rotation period. Both \textit{o}- and \textit{c}-filter data of those targets were folded with the period corresponding to strongest periodogram peak and visually inspected to ascertain the quality of the fold (for instance, we retained an extracted period if the periodogram-independent \textit{c}-filter data also folded commensurately with the \textit{o}-filter data). Using this methodology we report rotation periods for 16 L4 and 25 L5 Trojans of which 27 of those have previously reported periods in the Asteroid Light Curve Database\footnote{\url{http://alcdef.org}} \citep[LCDB;][Updated 2020 June 26]{warner2009}. Our extracted periods match 20 out of the 27 objects that had previously reported periods. Aliasing ambiguity in the extracted rotation period is a common problem with ATLAS data due to the diurnal cadence of the ATLAS observations which means that in most cases the ATLAS sampling frequency is lower than the rotation frequencies we are trying to resolve. In some cases in the ATLAS data, aliases of the actual rotation period can also have similar peak strengths in the periodograms and it becomes difficult to unambiguously extract a period. We extract the $\pm$2-, $\pm$1-, or $\pm$0.5-day alias periods of the strongest periodogram derived period for the 7 Trojans where our period did not match the literature period and found that in all cases the LCDB period was one of these aliases. This effect acts in both directions, and our derived period is also then an alias of the LCDB period. As these 7 objects have multiple listings of the same period in the LCDB and have been assigned a quality code $U=3$ (defined as an unambiguous period solution) we defer to the literature in these cases.  In Table \ref{table_rot_data} we report the 41 periods we could extract and also show the LCDB periods. Where our period does not match the LCDB period we indicate our best matching alias period and the corresponding alias window.

\subsection{Shape Distribution}

The L4 and L5 Trojan clouds were considered separately and shape distributions for each population were determined. Due to the unknown spin pole distribution of the Jupiter Trojans we assume both a case where all objects have spin poles perpendicular to the ecliptic and a case where all objects have spin pole latitudes $\theta = 50^{\circ}$. Although the real shape distribution of the population is highly dependent on the spin pole latitudes observed, any difference between the two clouds will remain regardless of the assumption used. Therefore, it is emphasised that although the mean values for elongation in the two clouds themselves may not be meaningful, any disparity between the L4 and L5 values is a real effect.

\begin{figure}
\begin{center}
\includegraphics[width=0.7\textwidth]{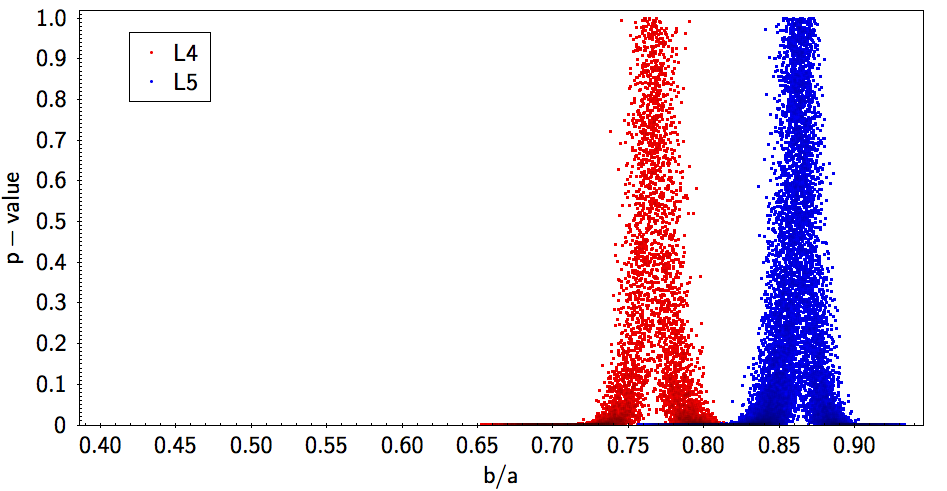}
\caption{Two-sample Kolmogorov-Smirnov test p-values from model populations compared to observed data for both L4 and L5 clouds as a function of axis ratio. Each of the 25,000
points show the KS statistic calculated by comparing the observed cumulative distribution function (CDF) to a model population with truncated Gaussian shape distributions. The axis ratio value corresponds to the median shape
for each model population.}
\label{ks_90}
\end{center}
\end{figure}

\begin{figure}
\begin{center}
\includegraphics[width=0.7\textwidth]{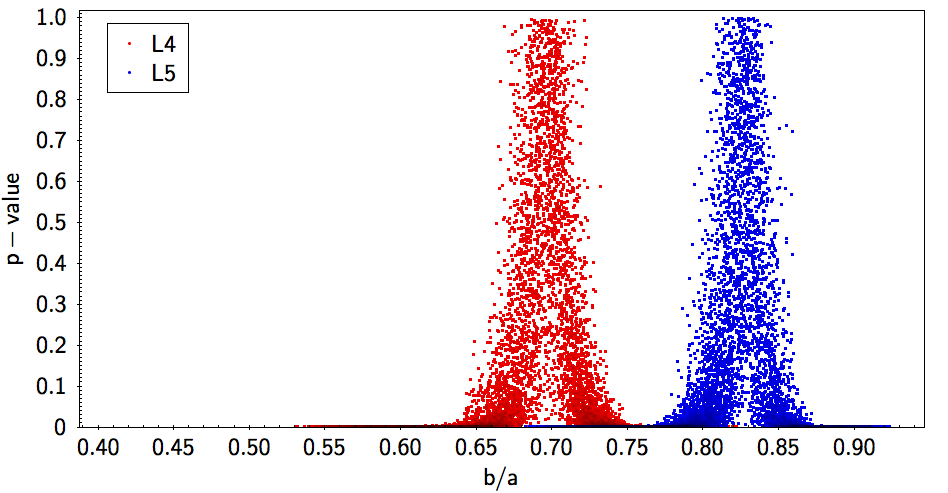}
\caption{The same information plotted as in Figure~\ref{ks_90} but now using the assumption of spin pole latitude $\theta=50^{\circ}$}
\label{ks_50}
\end{center}
\end{figure}

For the L4 cloud we obtain a best fit mean elongation of $\frac{b}{a} = 0.77 \pm 0.02$ and for L5 we obtain a value $\frac{b}{a} = 0.86 \pm 0.02$. Figure~\ref{ks_90} shows the p-value obtained from the KS test for a wide range of shape distributions for both L4 and L5 clouds plotted against the average elongation of the distribution. In this case the assumption of perpendicular spin poles is used. Figure~\ref{ks_50} shows the same information in the case of spin pole latitudes $\approx 50^{\circ}$. Potential mechanisms for this difference will be discussed in Section~\ref{subsec:explain}.

We considered the possibility of carrying out this modelling for individual families within the Trojans, however, the number of family members in this dataset is insufficient for this analysis. This represents a potential avenue for study when future surveys, e.g. LSST, come online.

\section{Discussion} \label{sec:discussion}

\subsection{Comparison of colours and phase curve parameters between L4 and L5}


\begin{figure*}
\begin{center}
\includegraphics[width=0.8\textwidth]{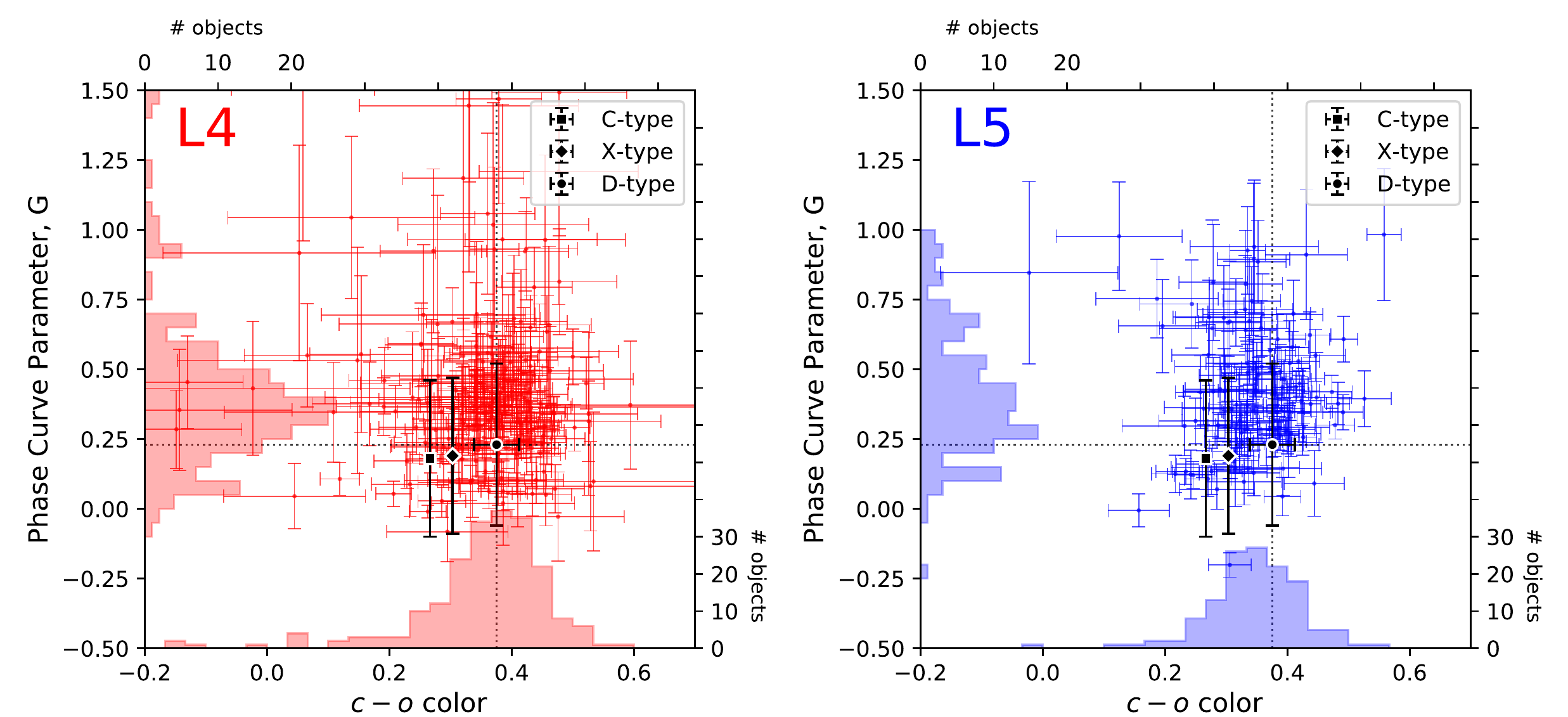}
	\caption{Scatter plots showing the derived phase curve parameters, G, and c-o colours for each of the L4 and L5 Trojan clouds. Using the K-S test suggests that we cannot rule out that the samples from the L4 and L5 clouds for G and colour were drawn from the same distribution.} 
	\label{L4_L5_color_rot}
\end{center}
\end{figure*}

In general, both the L4 and L5 objects have \textit{c-o} colours consistent with the expected D-type taxonomy's \textit{c-o} colour (see Figure \ref{L4_L5_color_rot}). However, the median colour of the L4 cloud appears at a slightly higher \textit{c-o} colour ($\sim 0.41$ mag) than that of the L5 cloud ($\sim 0.39$ mag) \sout{(compare the moving-average curves Figure \ref{L4_L5_color_rot})}. While the median colour of the L4 is slightly higher, the colour distribution of the L4 objects is much broader containing both redder objects and also objects with low C- or X-like \textit{c-o} colours (i.e. objects with less-red or even neutral slopes). Using a two-sample Kolmogorov-Smirnov test on the two distributions for suggests that the colours are not drawn from identical distributions but we cannot rule this out for slope parameter, G. (p-values: $0.01$ for c-o colour, 0.71 for slope parameter G)

\begin{figure}[!tbp]
  \centering
  \subfigure{\includegraphics[width=0.45\textwidth]{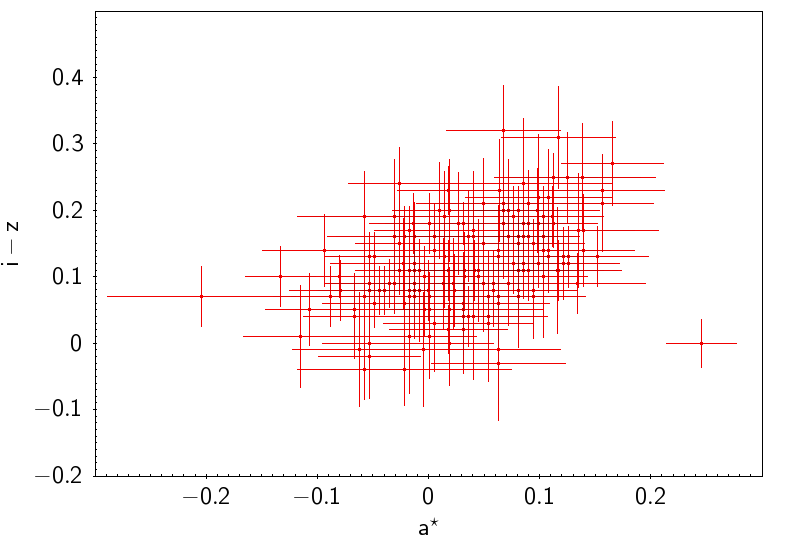}\label{fig:f1}}
  \subfigure{\includegraphics[width=0.45\textwidth]{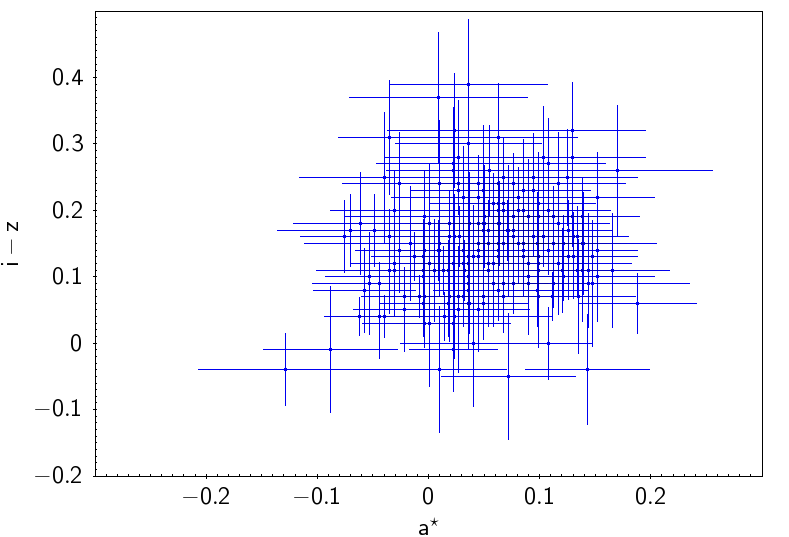}\label{fig:f2}}
  \subfigure{\includegraphics[width=0.45\textwidth]{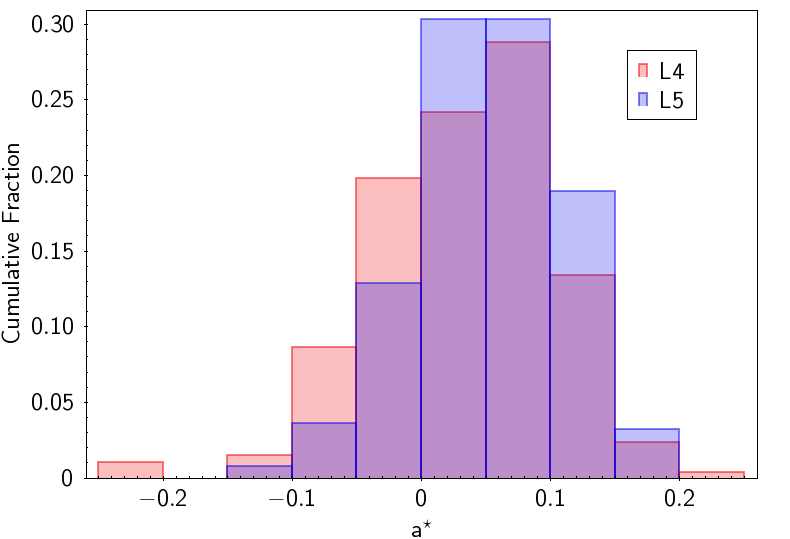}\label{fig:f3}}
  \subfigure{\includegraphics[width=0.45\textwidth]{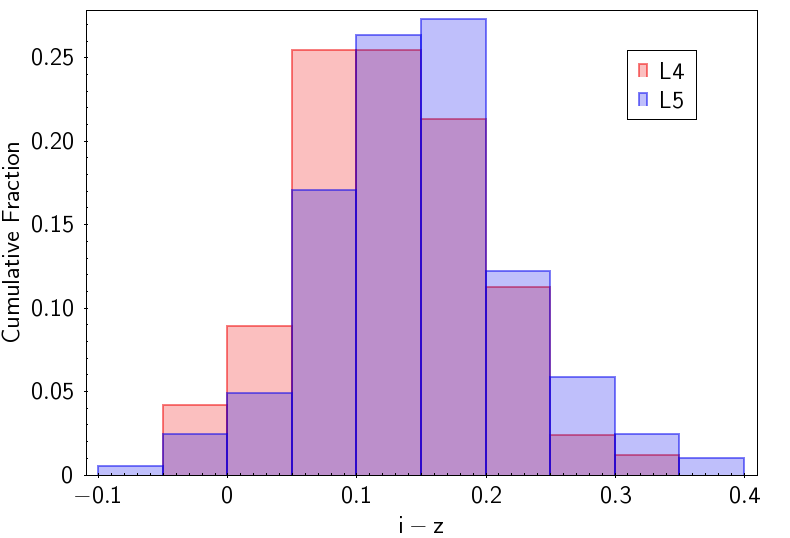}\label{fig:f4}}
  \caption{Four panels showing the i-z vs. a* colour distribution of the L4 and L5 Trojans as measured by the Sloan Digital Sky Survey. \textit{Top Left:} A colour-colour plot in i-z vs a* for the L4 Trojan cloud from SDSS. \textit{Top Right:} A colour-colour plot in i-z vs a* for the L5 Trojan cloud from SDSS. \textit{Bottom Left:} A histogram comparing the a* distributions of the L4 and L5 Trojan clouds. \textit{Bottom Right:} A histogram comparing the i-z distributions of the L4 and L5 Trojan clouds. From this we conclude that the colour distribution of the L4 and L5 clouds is slightly different in SDSS data as well as in ATLAS.}
  \label{mommertgif}
\end{figure}

We attemped to verify this colour discrepancy using data from the Sloan Digital Sky Survey (SDSS; \citealt{ivezic1998}). The colour-colour plots for the L4 and L5 Trojans from SDSS are given in Figure~\ref{mommertgif}. From these measurements, the L4 Trojans do show a broader distribution in a* than the L5 Trojans. Here, a* is a principal component used in asteroid colour analysis from SDSS as defined by \cite{ivezic2001}. Again in the case of the SDSS data, the KS test suggests that these two populations can not have been drawn from the same distribution. To assess whether this difference could be due to the presence of an abundance of Eurybates family objects in our sample, we cross-checked our list of targets with a list of family members and found only several objects in common.\\
\cite{szabo2007}, \cite{roig2008} and \cite{emery2011} all offer evidence of colour bimodality in the Trojan population, yet none draw clear conclusions about differences in colour abundances between the two clouds. \cite{szabo2007} find a difference in colour distribution between L4 and L5, however, this is removed through normalisation due to the different number densities of two clouds. \cite{roig2008} observe a colour bimodality, and see a difference in the distributions of each cloud but removing family members from their analysis they find the two clouds to be identical. As only a few Eurybates family members are present in our data, they alone cannot be producing this effect. Figure~\ref{mommertgif} shows the i-z and a* colours for all objects shared in our ATLAS sample and SDSS archival data. The bimodality is not readily visible in these plots, however, limiting this to only the brightest targets ($H < 12$) does show this effect. This is due to large uncertainties on objects on Trojans where $H > 13$.
We do not observe colour bimodality in ATLAS c-o measurements as the wavelength ranges here span the 'kink' in the spectra of 'less-red' Trojans (\citealt{emery2011}). This has the effect of making the expected c-o colours for each group closer together, preventing the bimodality from being observed.



\subsection{Potential explanations for the difference in apparent elongation between L4 and L5}\label{subsec:explain}

We consider a range of mechanisms which may produce the difference in shape distribution between the two Trojan clouds and assess their validity, from least likely to most likely.

\subsubsection{A difference in spin-pole distribution}

As previously stated the assumption of the spin-pole distribution of the population is important in obtaining its overall shape distribution. In the investigation we have assumed that both clouds have identical spin-pole distributions. However, if this is not the case, a difference in spin-pole distribution could be invoked to explain the difference in shape distribution between the two clouds.
Assuming a spin-pole distribution where the poles are aligned toward the observers produces a shape distribution that appears more spherical,
while a more randomly oriented distribution 
more accurately reveals an underlying elongated shape distribution.

\begin{figure}
\begin{center}
\includegraphics[width=0.7\textwidth]{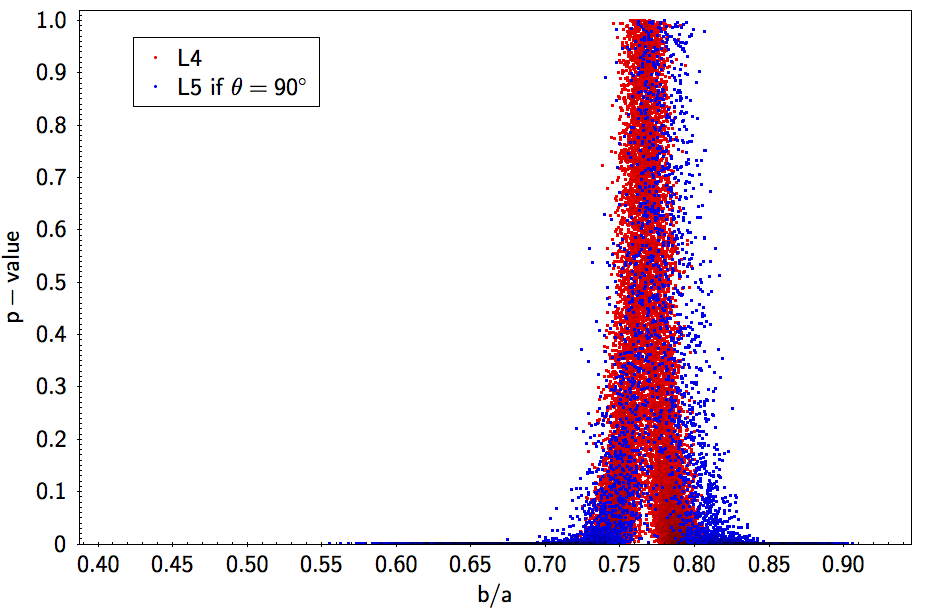}
\caption{KS statistics from model populations compared to observed data for the L4 (assuming $\theta = 0^{\circ}$) and L5 assuming $\theta = 90^{\circ}$ clouds as a function of axis ratio. The axis ratio value corresponds to the median shape
for each model population.}
\label{fig_90}
\end{center}
\end{figure}

Keeping the spin-pole distribution of the more elongated L4 cloud constant we vary the spin-pole distributions of the L5 cloud in order to try to produce identical results. This is presented in Figure~\ref{fig_90}. We find that in order to produce the same best fit shape distribution the required spin-pole distribution of the more elongated L4 cloud is for all objects to have spin-pole latitudes parallel to the ecliptic while the L5 cloud objects are all aligned to spin-poles perpendicular to the ecliptic. 
We reject this solution as
too unlikely. Although a difference in spin-pole distribution cannot by itself explain the difference in shape distribution, it is possible that it may be a contributing factor along with a stronger mechanism and as such we can not rule out a difference in the spin pole distributions of the two clouds.

\subsubsection{An abundance of slow rotators}\label{subsubsec:slow}

The spin-rate distribution is also a key component of the shape distribution model. If the spin-rate distribution assumed under-represents slow rotating asteroids then partial lightcurves for these objects may be mistaken for much lower amplitude objects with a shorter rotation period. For example, if the model spin-rate distribution allows a maximum of $10\%$ of the population to be slow rotators and the real abundance is $20\%$ then the overflow, i.e. half of the slow rotators in the population, will be treated as low-amplitude average rotators. A significant proportion of these objects could result in a best-fit shape distribution skewed toward more spherical shapes than are really present in the population.

\begin{figure}
\begin{center}
\includegraphics[width=0.7\textwidth]{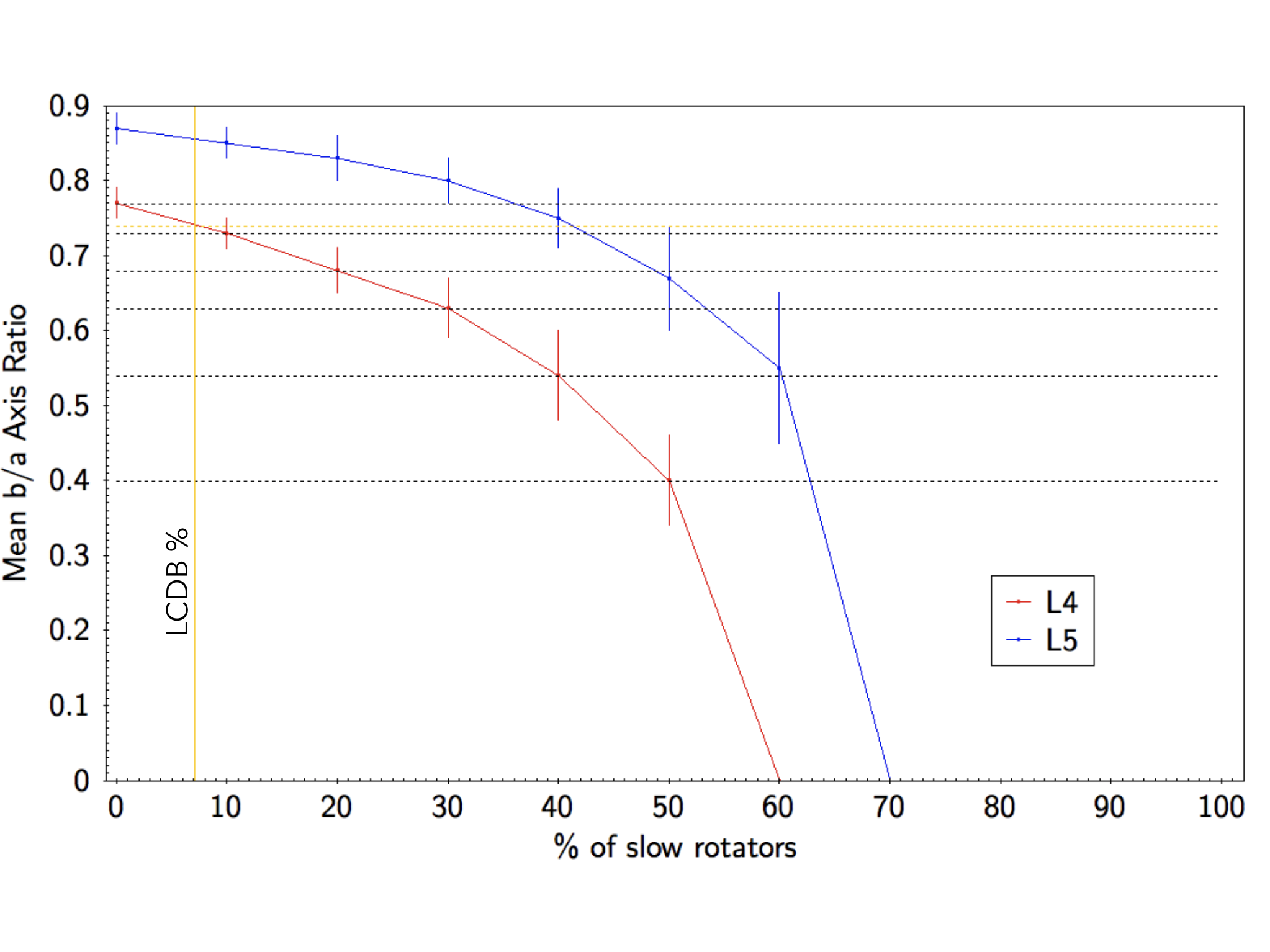}
\caption{The mean axis ratio for the L4 (red) and L5 (blue) shape distributions with a varying proportion of slow rotators (P$> 100$ h) injected into the model. The horizontal dashed lines correspond to the values obtained for the L4 cloud and the intercept of this line with the L5 curve is the required proportion of slow rotators to give the same axis ratio. The vertical yellow line corresponds to the proportion of Jupiter Trojans with $P > 100$ h as recorded in the Light Curve Database (LCDB; \citealt{warner2009}). The tend toward zero beyond $50\%$ for L4 and $60\%$ for L5 represents the point at which no model fits apply and the curves should not be believed beyond these points.}
\label{slowrot}
\end{center}
\end{figure}

Assuming the more elongated L4 cloud to have a fixed spin-rate distribution we vary the distribution for the L5 cloud by artificially injecting a proportion of slow rotating objects ($P > 100$ h) into the model and obtaining the best fit shape distribution in each case. Figure~\ref{slowrot} shows how the derived shape distribution of the L5 cloud varies depending on the proportion of slow rotators assumed. Here, the spin pole distribution of both clouds is kept constant. A excess in the proportion of slow rotators alone of $30-40\%$ is required in the L5 cloud to bring both clouds to the same mean axis ratio. A proportion of slow rotators has been identified among Jupiter Trojans, with estimates from \cite{szabo2017} and \cite{ryan2017} suggesting a proportion of slow rotators ($P > 100$ h) of $\sim 15\%$. However, there has been no evidence of a disparity in this proportion between the two Trojan clouds (\citealt{french2015}). As this is a relatively large difference in the proportion of slow rotators we consider this to be unlikely. Previous work using the Transiting Exoplanet Satellite Survey (TESS) has shown a proportion of slow rotators in the main asteroid belt, though one much smaller than that required here (\citealt{mcneill2020}).

\subsubsection{Collisional effects}\label{subsec:coll}

If the difference in shape shown in the two clouds is a real difference in shape and not a function of differing rotational properties (spin-pole orientations, rotation periods)
between the two clouds, this may imply a different collisional evolution within each cloud. The Trojan clouds are often considered to be a collisionless environment, but this is generally in reference to external collisions, i.e., collisions between a Trojan and an object from outside the population. Trojans can undergo collisions with Hilda objects, however, Trojan-Trojan collisions will dominate interactions. This is backed up by the presence of a collisional family in the L4 Trojan cloud dynamically linked to (3548) Eurybates (\citealt{broz2011}). These Eurybates objects are likely to be C-type objects, however, there is only a very small proportion of these objects in our data set and these can be easily excluded.

The L4 cloud contains more objects than the L5, so it therefore follows that collisional interactions in this cloud will be more frequent. This difference may produce a systematic difference in shape distribution between the two populations. 
From \cite{delloro1998} we have 
equation~\ref{eqn:tau},
which gives the expected timescale over which sub-catastrophic collisions occur in each of the two Trojan clouds where $P_{i}$ is the intrinsic collision probability of each cloud, $R$ is the radius of the target objects, and $N(r)$ is the number of objects with a radius greater than $r$:

\begin{equation}\label{eqn:tau}
    \tau=\frac{1}{P_{i}R^{2}N(r)}.
\end{equation}

\noindent
We calculate $\tau$ for 
projectile radii $r$ larger than 1~km, corresponding to the lower completeness limit of the Trojan size distribution for each Trojan cloud as determined by \cite{yoshida2008}.

\begin{figure}
\begin{center}
\includegraphics[width=0.7\textwidth]{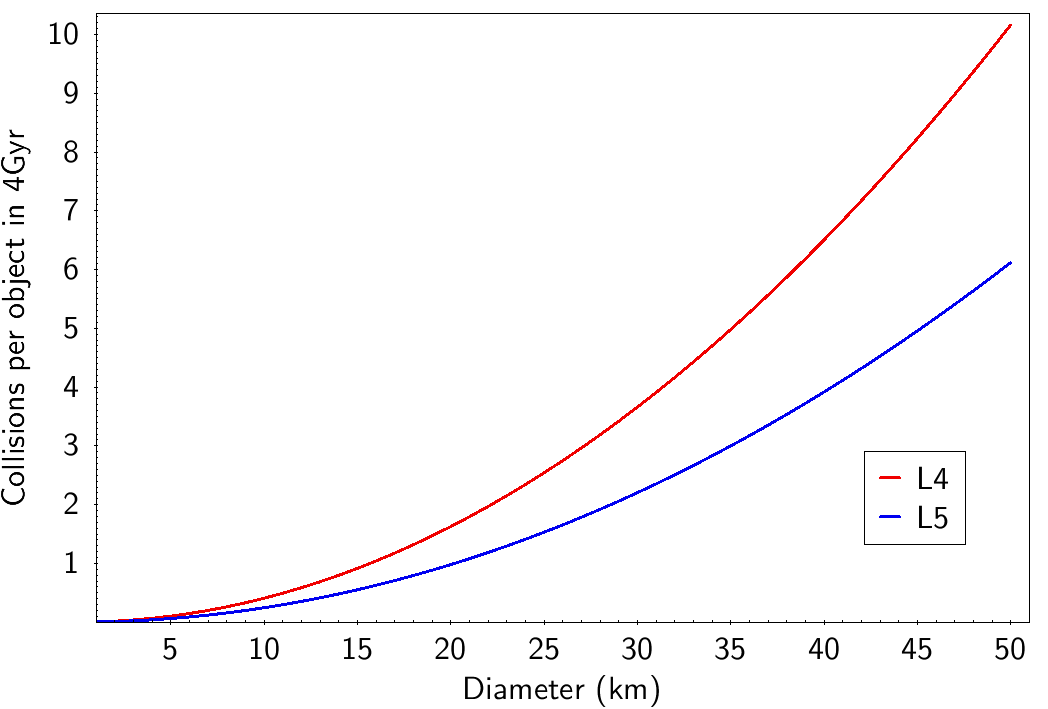}
\caption{The expected abundances of collision between a projectile $D > 2$ km and a target object of varying size in both the L4 and L5 Trojan clouds in a period of 4Gyr.}
\label{fig:coll}
\end{center}
\end{figure}

We assume average impact velocities of $5.06$ and $4.96$ km s$^{-1}$ for the L4 and L5 clouds respectively (\citealt{delloro1998}). We consider the case of a $D > 2$ km
projectile colliding with a target of variable size. In each case we calculate the Trojan-Trojan collisional timescale for this object, and hence the number of collisions it would experience in 4Gyr, if it were situated in both the L4 and L5 clouds. This is shown in Figure~\ref{fig:coll}.

The collisional timescale of objects in the L4 cloud is shorter than the L5 regardless of the size of the target simply due to the increased number density of the population. Since the L4 cloud is more collisionally evolved we conclude that if collisions are responsible for the shape difference between the two clouds then sub-catastrophic collisions must make a population more elongated over time. Both \cite{domokos2009} and \cite{henych2015} simulated the effect of
subcatastrophic collisions on the elongation of small asteroids ($D < 20$ km). They
demonstrated that the cumulative effect of collisions should lead to an increase in the
target object’s elongation, occurring over shorter timescales at smaller sizes. However,
it is worth noting that \cite{henych2015} found that the estimated timescales for this process to occur
are significantly longer than the collisional disruption timescales for the asteroids in
question, a discrepancy which is not fully explained by the longer collisional timescales of objects in the Trojan clouds. \cite{henych2015} simulate the time taken for an object to go from a 2:1 axis ratio to a 3:1 axis ratio due to erosion from sub-catastrophic collisions. For a 10 km object this is found to be of order $1$ Gyr, a longer timescale than the collisional lifetime of such an object in the main belt. Due to the relatively lower number density of the Trojan clouds compared to the main belt, the collisional lifetime in this region will be longer. However, until a similar modelling work of sub-catastrophic collisional erosion is carried out for the Trojan population it is impossible to reliably compare these timescales.

Work by \cite{wong2014} and \cite{wong2015} discovered a dependence of the abundances of 'less-red' Trojans and 'red' Trojans with diameter. This effect showed that there were a greater proportion of 'less-red' objects at small sizes. This trend is hypothesised to be collisional in nature. \cite{wong2014} assume both Trojan sub-groups to have similar interior composition, catastrophic collisions of each type of object will produce spectroscopically identical resultant bodies which will resemble 'less-red' objects. On a sufficient timescale this leads to the increasing abundance of less-red objects and the depletion of red objects. If the colour difference observed is due to only the surface of 'red' and 'less-red' objects, due to e.g. space weathering, it follows that sub-catastrophic collisions will also potentially produce a similar effect by exposing underlying material. The collisions considered in this paper are limited to impactors of $D > 2$ km due to the completeness of the known size distribution for Trojans. Smaller objects will contribute to this effect and when surveys carry out further observations of small Trojans to improve this size distribution this will represent an interesting avenue for future work in collisional modelling.



\section{Conclusions} \label{sec:conclusions}

Using data from the ATLAS survey, we derive phase functions and $c-o$ colours for 266 Trojans. The colours obtained here from ATLAS show different distributions of colours for the L4 cloud than the L5.
We also present shape distributions derived for each of the Trojan clouds. The L4 population appears to show a more elongated shape distribution than the L5 cloud. We rule out that this difference could be solely a result of different spin-pole distributions between the two clouds.
We also rule out a difference in the abundance of slow rotators (P $> 100$ h) between the two clouds as 
unrealistic.
This leaves as the most plausible explanation the different collisional environments in the two clouds. The collisional timescale in the L4 cloud is shorter, and therefore this cloud is more collisionally evolved. We conclude that collisions may have made this population more elongated on average. This shape discrepancy should be confirmed using observations from The Rubin Observatory Legacy Survey of Space and Time (LSST) which will obtain observations for a larger population of Jupiter Trojans (approximately 280,000 over the lifetime of the survey; \citealt{lsst2009}). Furthermore, comparisons of the apparent cratering records for L4 and L5 targets as measured by the Lucy mission will provide a good test of this proposed hypothesis.




\acknowledgments
\textbf{Acknowledgments}\\
We thank the anonymous referees for their input which has led to significant improvement of this manuscript. 
A.M. was supported by the Arizona Board of Regents' Regent Innovation Fund. This work has made use of data from the Asteroid Terrestrial-impact Last Alert System (ATLAS) project. ATLAS is primarily funded to search for near-Earth asteroids through NASA grants NN12AR55G, 80NSSC18K0284, and 80NSSC18K1575; byproducts of the NEO search include images and catalogs from the survey area. The ATLAS science products have been made possible through the contributions of the University of Hawaii Institute for Astronomy, the Queen's University Belfast, the Space Telescope Science Institute, and the South African Astronomical Observatory. The authors thank Bill Bottke for discussions which improved this manuscript.
\\

%

\vspace{5mm}
\facilities{ATLAS (\citealt{Tonry2018})}


\software{astropy \citep{astropy2013}}



\appendix

\startlongtable
\begin{deluxetable*}{|c|l|c|r|r|r|}
	\tablecaption{ATLAS colours and phase curve parameters \label{table_col_data}}
	\tablehead{
		\colhead{No.} & \colhead{Object Name} & \colhead{Trojan Group} & \colhead{\textit{H}\tablenotemark{a} } & \colhead{\textit{c-o} color}& \colhead{G-parameter} \\
		\colhead{} & \colhead{} & \colhead{} & \colhead{(mag)}& \colhead{(mag)}& \colhead{}
	}
	\startdata
	001&10247 Amphiaraos (6629 P-L)&L4&11.2&0.248$\pm$0.065&0.392$\pm$0.070\\
002&10664 Phemios (5187 T-2)&L4&11.3&-0.148$\pm$0.107&0.285$\pm$0.140\\
003&10989 Dolios (1973 SL1)&L4&11.3&0.320$\pm$0.064&0.319$\pm$0.105\\
004&11252 Laertes (1973 SA2)&L4&10.7&0.387$\pm$0.050&0.482$\pm$0.065\\
005&11351 Leucus (1997 TS25)&L4&10.9&0.355$\pm$0.078&0.255$\pm$0.184\\
006&11395 (1998 XN77)&L4&9.9&0.207$\pm$0.029&0.054$\pm$0.045\\
007&11396 (1998 XZ77)&L4&10.6&0.368$\pm$0.043&0.330$\pm$0.079\\
008&11397 (1998 XX93)&L4&10.2&0.303$\pm$0.042&0.670$\pm$0.122\\
009&11429 Demodokus (4655 P-L)&L4&10.4&0.471$\pm$0.048&0.072$\pm$0.087\\
010&1143 Odysseus (1930 BH)&L4&8.3&0.385$\pm$0.013&0.298$\pm$0.040\\
011&114694 (2003 FC99)&L4&11.8&0.372$\pm$0.040&0.930$\pm$0.295\\
012&12238 Actor (1987 YU1)&L4&10.9&0.276$\pm$0.050&0.283$\pm$0.104\\
013&12658 Peiraios (1973 SL)&L4&11.2&-0.130$\pm$0.091&0.454$\pm$0.166\\
014&12714 Alkimos (1991 GX1)&L4&10.2&0.425$\pm$0.039&0.387$\pm$0.103\\
015&12916 Eteoneus (1998 TL15)&L4&11.4&0.394$\pm$0.050&0.456$\pm$0.122\\
016&12917 (1998 TG16)&L4&11.4&0.138$\pm$0.202&1.044$\pm$0.291\\
017&12921 (1998 WZ5)&L4&11.0&0.252$\pm$0.059&0.588$\pm$0.150\\
018&12973 Melanthios (1973 SY1)&L4&11.4&0.289$\pm$0.087&0.222$\pm$0.146\\
019&12974 Halitherses (1973 SB2)&L4&11.3&0.336$\pm$0.073&0.103$\pm$0.134\\
020&13060 (1991 EJ)&L4&10.8&0.461$\pm$0.069&0.659$\pm$0.282\\
021&13062 Podarkes (1991 HN)&L4&11.2&0.422$\pm$0.071&0.923$\pm$0.143\\
022&13182 (1996 SO8)&L4&10.7&0.259$\pm$0.051&0.238$\pm$0.077\\
023&13183 (1996 TW)&L4&10.7&0.369$\pm$0.035&0.472$\pm$0.150\\
024&13184 Augeias (1996 TS49)&L4&11.1&0.416$\pm$0.076&0.411$\pm$0.147\\
025&13323 (1998 SQ)&L4&11.1&0.342$\pm$0.069&0.393$\pm$0.101\\
026&13331 (1998 SU52)&L4&11.4&0.594$\pm$0.120&0.372$\pm$0.229\\
027&13362 (1998 UQ16)&L4&11.0&0.409$\pm$0.035&0.310$\pm$0.097\\
028&13366 (1998 US24)&L4&11.2&-0.023$\pm$0.257&0.432$\pm$0.240\\
029&13372 (1998 VU6)&L4&11.3&0.432$\pm$0.065&0.363$\pm$0.135\\
030&13383 (1998 XS31)&L4&11.3&0.295$\pm$0.099&-0.082$\pm$0.108\\
031&13385 (1998 XO79)&L4&10.9&0.411$\pm$0.027&0.497$\pm$0.093\\
032&13463 Antiphos (5159 T-2)&L4&11.2&0.349$\pm$0.090&0.451$\pm$0.233\\
033&13694 (1997 WW7)&L4&10.9&0.191$\pm$0.077&0.366$\pm$0.104\\
034&13782 (1998 UM18)&L4&11.6&0.279$\pm$0.161&0.662$\pm$0.462\\
035&1404 Ajax (1936 QW)&L4&9.3&0.386$\pm$0.036&0.063$\pm$0.068\\
036&14268 (2000 AK156)&L4&10.5&0.478$\pm$0.094&0.814$\pm$0.190\\
037&1437 Diomedes (1937 PB)&L4&8.2&0.294$\pm$0.017&0.194$\pm$0.024\\
038&14690 (2000 AR25)&L4&10.6&0.423$\pm$0.053&0.533$\pm$0.118\\
039&14707 (2000 CC20)&L4&11.3&0.252$\pm$0.054&0.592$\pm$0.126\\
040&15033 (1998 VY29)&L4&10.7&0.109$\pm$0.179&0.346$\pm$0.179\\
041&15398 (1997 UZ23)&L4&10.9&0.433$\pm$0.032&0.218$\pm$0.139\\
042&15436 (1998 VU30)&L4&9.1&0.358$\pm$0.032&0.345$\pm$0.049\\
043&15440 (1998 WX4)&L4&9.6&0.469$\pm$0.018&0.344$\pm$0.060\\
044&15521 (1999 XH133)&L4&11.2&0.361$\pm$0.077&1.058$\pm$0.288\\
045&15527 (1999 YY2)&L4&10.9&0.383$\pm$0.038&0.110$\pm$0.068\\
046&15529 (2000 AA80)&L4&11.4&0.272$\pm$0.087&0.924$\pm$0.294\\
047&15535 (2000 AT177)&L4&10.6&0.378$\pm$0.063&0.090$\pm$0.078\\
048&15536 (2000 AG191)&L4&11.3&0.148$\pm$0.294&0.532$\pm$0.406\\
049&15539 (2000 CN3)&L4&10.5&0.431$\pm$0.026&0.650$\pm$0.116\\
050&15651 Tlepolemos (9612 P-L)&L4&11.3&0.343$\pm$0.105&0.457$\pm$0.213\\
051&1583 Antilochus (1950 SA)&L4&8.6&0.357$\pm$0.015&0.134$\pm$0.050\\
052&16099 (1999 VQ24)&L4&10.7&0.340$\pm$0.038&0.431$\pm$0.068\\
053&1647 Menelaus (1957 MK)&L4&10.5&0.452$\pm$0.037&0.261$\pm$0.058\\
054&16974 Iphthime (1998 WR21)&L4&9.9&0.400$\pm$0.023&0.419$\pm$0.044\\
055&17351 Pheidippos (1973 SV)&L4&11.3&0.385$\pm$0.155&0.966$\pm$0.484\\
056&1749 Telamon (1949 SB)&L4&9.5&0.419$\pm$0.030&0.242$\pm$0.062\\
057&17874 (1998 YM3)&L4&11.7&0.437$\pm$0.062&0.794$\pm$0.144\\
058&18060 (1999 XJ156)&L4&11.1&0.319$\pm$0.092&0.178$\pm$0.088\\
059&18062 (1999 XY187)&L4&10.9&0.465$\pm$0.055&0.251$\pm$0.119\\
060&18063 (1999 XW211)&L4&11.2&0.404$\pm$0.043&0.682$\pm$0.227\\
061&18071 (2000 BA27)&L4&11.6&0.314$\pm$0.110&0.228$\pm$0.173\\
062&18263 Anchialos (5167 T-2)&L4&11.4&0.334$\pm$0.080&0.479$\pm$0.123\\
063&1868 Thersites (2008 P-L)&L4&9.5&0.421$\pm$0.021&0.398$\pm$0.047\\
064&1869 Philoctetes (4596 P-L)&L4&11.1&0.229$\pm$0.054&0.171$\pm$0.098\\
065&19725 (1999 WT4)&L4&10.8&0.345$\pm$0.038&0.364$\pm$0.085\\
066&19913 Aigyptios (1973 SU1)&L4&11.2&0.345$\pm$0.067&0.210$\pm$0.092\\
067&20144 (1996 RA33)&L4&11.4&0.238$\pm$0.143&0.399$\pm$0.235\\
068&20424 (1998 VF30)&L4&10.6&0.423$\pm$0.107&0.317$\pm$0.134\\
069&20729 (1999 XS143)&L4&10.5&0.390$\pm$0.060&0.496$\pm$0.126\\
070&20738 (1999 XG191)&L4&11.6&0.343$\pm$0.095&0.698$\pm$0.260\\
071&21284 Pandion (1996 TC51)&L4&11.5&0.458$\pm$0.069&0.635$\pm$0.129\\
072&21372 (1997 TM28)&L4&11.8&0.059$\pm$0.117&1.505$\pm$0.544\\
073&2146 Stentor (1976 UQ)&L4&10.0&0.391$\pm$0.024&0.507$\pm$0.168\\
074&2148 Epeios (1976 UW)&L4&10.8&0.377$\pm$0.034&0.343$\pm$0.091\\
075&21595 (1998 WJ5)&L4&10.7&0.380$\pm$0.028&0.199$\pm$0.111\\
076&21599 (1998 WA15)&L4&11.4&0.053$\pm$0.223&0.917$\pm$0.387\\
077&21601 (1998 XO89)&L4&10.1&0.309$\pm$0.031&0.425$\pm$0.069\\
078&21900 Orus (1999 VQ10)&L4&10.0&0.364$\pm$0.038&0.365$\pm$0.059\\
079&22008 (1999 XM71)&L4&11.6&0.280$\pm$0.093&0.476$\pm$0.135\\
080&22014 (1999 XQ96)&L4&10.2&0.463$\pm$0.087&0.495$\pm$0.158\\
081&22052 (2000 AQ14)&L4&11.2&0.321$\pm$0.081&0.305$\pm$0.103\\
082&22054 (2000 AP21)&L4&11.3&-0.143$\pm$0.184&0.354$\pm$0.217\\
083&22055 (2000 AS25)&L4&11.1&0.430$\pm$0.037&0.430$\pm$0.120\\
084&22059 (2000 AD75)&L4&11.2&0.379$\pm$0.036&0.445$\pm$0.162\\
085&22149 (2000 WD49)&L4&10.3&0.388$\pm$0.046&0.462$\pm$0.124\\
086&22203 Prothoenor (6020 P-L)&L4&11.6&0.313$\pm$0.034&0.231$\pm$0.049\\
087&2260 Neoptolemus (1975 WM1)&L4&9.3&0.380$\pm$0.020&0.353$\pm$0.057\\
088&23075 (1999 XV83)&L4&11.0&0.066$\pm$0.103&0.550$\pm$0.185\\
089&23126 (2000 AK95)&L4&11.7&0.448$\pm$0.051&0.374$\pm$0.081\\
090&23135 (2000 AN146)&L4&10.0&0.286$\pm$0.039&0.028$\pm$0.058\\
091&23269 (2000 YH62)&L4&11.5&0.458$\pm$0.265&0.365$\pm$0.208\\
092&23285 (2000 YH119)&L4&11.1&0.154$\pm$0.084&0.554$\pm$0.281\\
093&23480 (1991 EL)&L4&11.3&0.478$\pm$0.110&1.493$\pm$0.514\\
094&23622 (1996 RW29)&L4&11.4&0.333$\pm$0.108&0.101$\pm$0.146\\
095&23709 (1997 TA28)&L4&11.6&0.306$\pm$0.137&0.285$\pm$0.154\\
096&23958 (1998 VD30)&L4&10.2&0.378$\pm$0.054&0.296$\pm$0.100\\
097&23970 (1998 YP6)&L4&11.0&0.476$\pm$0.108&-0.028$\pm$0.160\\
098&24244 (1999 XY101)&L4&10.9&0.424$\pm$0.084&0.932$\pm$0.183\\
099&24275 (1999 XW167)&L4&11.2&0.446$\pm$0.129&0.564$\pm$0.170\\
100&24312 (1999 YO22)&L4&11.6&0.526$\pm$0.118&0.314$\pm$0.144\\
101&24313 (1999 YR27)&L4&10.8&0.425$\pm$0.029&0.229$\pm$0.107\\
102&24390 (2000 AD177)&L4&11.6&0.477$\pm$0.130&1.209$\pm$0.477\\
103&24403 (2000 AX193)&L4&11.3&0.330$\pm$0.179&1.445$\pm$0.596\\
104&24485 (2000 YL102)&L4&11.4&0.366$\pm$0.042&0.367$\pm$0.114\\
105&24486 (2000 YR102)&L4&11.1&0.338$\pm$0.031&0.349$\pm$0.068\\
106&24505 (2001 BZ)&L4&11.1&0.391$\pm$0.086&0.217$\pm$0.154\\
107&24506 (2001 BS15)&L4&10.7&0.431$\pm$0.051&0.311$\pm$0.145\\
108&24534 (2001 CX27)&L4&11.3&0.319$\pm$0.046&0.341$\pm$0.166\\
109&24537 (2001 CB35)&L4&11.1&0.526$\pm$0.080&0.340$\pm$0.292\\
110&2456 Palamedes (1966 BA1)&L4&9.2&0.389$\pm$0.039&0.375$\pm$0.089\\
111&24587 Kapaneus (4613 T-2)&L4&11.5&0.332$\pm$0.095&1.921$\pm$0.981\\
112&25895 (2000 XN9)&L4&11.1&0.410$\pm$0.047&0.100$\pm$0.165\\
113&25911 (2001 BC76)&L4&11.7&0.321$\pm$0.099&1.185$\pm$0.544\\
114&2759 Idomeneus (1980 GC)&L4&10.0&0.344$\pm$0.025&0.400$\pm$0.099\\
115&2797 Teucer (1981 LK)&L4&8.7&0.470$\pm$0.016&0.280$\pm$0.091\\
116&2920 Automedon (1981 JR)&L4&8.7&0.437$\pm$0.022&0.479$\pm$0.054\\
117&30102 (2000 FC1)&L4&10.8&0.358$\pm$0.050&0.235$\pm$0.137\\
118&3063 Makhaon (1983 PV)&L4&8.5&0.407$\pm$0.016&0.416$\pm$0.048\\
119&31835 (2000 BK16)&L4&11.5&0.342$\pm$0.062&0.583$\pm$0.227\\
120&3391 Sinon (1977 DD3)&L4&10.3&0.395$\pm$0.080&0.560$\pm$0.207\\
121&3548 Eurybates (1973 SO)&L4&9.8&0.333$\pm$0.029&0.093$\pm$0.063\\
122&3564 Talthybius (1985 TC1)&L4&9.4&0.452$\pm$0.038&0.237$\pm$0.055\\
123&35673 (1998 VQ15)&L4&11.6&0.045$\pm$0.116&0.045$\pm$0.117\\
124&3596 Meriones (1985 VO)&L4&9.3&0.405$\pm$0.034&0.247$\pm$0.118\\
125&36259 (1999 XM74)&L4&11.4&0.382$\pm$0.074&0.419$\pm$0.188\\
126&36267 (1999 XB211)&L4&10.8&0.389$\pm$0.056&0.344$\pm$0.124\\
127&3709 Polypoites (1985 TL3)&L4&9.1&0.343$\pm$0.018&0.398$\pm$0.054\\
128&37297 (2001 BQ77)&L4&11.6&0.370$\pm$0.156&1.018$\pm$0.438\\
129&37298 (2001 BU80)&L4&11.7&0.534$\pm$0.255&0.098$\pm$0.249\\
130&37714 (1996 RK29)&L4&12.2&0.340$\pm$0.089&0.367$\pm$0.169\\
131&3793 Leonteus (1985 TE3)&L4&8.8&0.119$\pm$0.032&0.107$\pm$0.059\\
132&3794 Sthenelos (1985 TF3)&L4&10.5&0.394$\pm$0.032&0.056$\pm$0.058\\
133&3801 Thrasymedes (1985 VS)&L4&11.1&0.456$\pm$0.042&0.050$\pm$0.111\\
134&38050 (1998 VR38)&L4&9.9&0.362$\pm$0.021&0.314$\pm$0.064\\
135&38607 (2000 AN6)&L4&11.7&0.168$\pm$0.094&0.376$\pm$0.150\\
136&38610 (2000 AU45)&L4&11.5&0.455$\pm$0.131&0.964$\pm$0.304\\
137&39264 (2000 YQ139)&L4&10.8&0.398$\pm$0.028&0.525$\pm$0.128\\
138&4007 Euryalos (1973 SR)&L4&10.3&0.447$\pm$0.041&0.199$\pm$0.098\\
139&4035 (1986 WD)&L4&9.6&0.437$\pm$0.019&0.287$\pm$0.059\\
140&4057 Demophon (1985 TQ)&L4&10.1&0.374$\pm$0.067&0.324$\pm$0.090\\
141&4060 Deipylos (1987 YT1)&L4&9.3&0.268$\pm$0.019&0.181$\pm$0.052\\
142&4063 Euforbo (1989 CG2)&L4&8.7&0.395$\pm$0.020&0.405$\pm$0.053\\
143&4068 Menestheus (1973 SW)&L4&9.5&0.385$\pm$0.029&0.287$\pm$0.058\\
144&4086 Podalirius (1985 VK2)&L4&9.2&0.319$\pm$0.069&0.374$\pm$0.146\\
145&41379 (2000 AS105)&L4&11.4&0.269$\pm$0.116&0.450$\pm$0.229\\
146&4138 Kalchas (1973 SM)&L4&10.1&0.301$\pm$0.054&0.095$\pm$0.069\\
147&42168 (2001 CT13)&L4&11.6&0.386$\pm$0.117&0.020$\pm$0.151\\
148&42367 (2002 CQ134)&L4&11.0&0.379$\pm$0.070&1.469$\pm$0.376\\
149&42554 (1996 RJ28)&L4&11.7&0.408$\pm$0.072&0.196$\pm$0.145\\
150&4489 (1988 AK)&L4&9.0&0.450$\pm$0.021&0.232$\pm$0.043\\
151&4501 Eurypylos (1989 CJ3)&L4&10.5&0.319$\pm$0.051&0.217$\pm$0.108\\
152&4543 Phoinix (1989 CQ1)&L4&9.7&0.368$\pm$0.064&0.555$\pm$0.114\\
153&46676 (1996 RF29)&L4&12.0&0.304$\pm$0.054&0.449$\pm$0.088\\
154&4833 Meges (1989 AL2)&L4&9.0&0.378$\pm$0.038&0.195$\pm$0.060\\
155&4834 Thoas (1989 AM2)&L4&9.1&0.468$\pm$0.020&0.349$\pm$0.052\\
156&4835 (1989 BQ)&L4&10.6&0.415$\pm$0.076&0.671$\pm$0.186\\
157&4836 Medon (1989 CK1)&L4&9.5&0.454$\pm$0.023&0.298$\pm$0.062\\
158&4902 Thessandrus (1989 AN2)&L4&9.9&0.427$\pm$0.040&0.277$\pm$0.070\\
159&4946 Askalaphus (1988 BW1)&L4&10.2&0.522$\pm$0.029&0.450$\pm$0.092\\
160&5012 Eurymedon (9507 P-L)&L4&10.6&0.338$\pm$0.032&0.368$\pm$0.069\\
161&5023 Agapenor (1985 TG3)&L4&10.4&0.320$\pm$0.027&0.307$\pm$0.080\\
162&5025 (1986 TS6)&L4&10.4&0.431$\pm$0.029&0.511$\pm$0.126\\
163&5027 Androgeos (1988 BX1)&L4&9.7&0.433$\pm$0.034&0.284$\pm$0.048\\
164&5028 Halaesus (1988 BY1)&L4&10.3&0.415$\pm$0.048&0.305$\pm$0.124\\
165&5041 Theotes (1973 SW1)&L4&10.7&0.331$\pm$0.072&0.402$\pm$0.108\\
166&5123 (1989 BL)&L4&10.0&0.363$\pm$0.047&0.063$\pm$0.063\\
167&5126 Achaemenides (1989 CH2)&L4&10.6&0.210$\pm$0.058&0.349$\pm$0.094\\
168&5209 (1989 CW1)&L4&10.3&0.377$\pm$0.034&0.464$\pm$0.074\\
169&5244 Amphilochos (1973 SQ1)&L4&10.5&0.297$\pm$0.050&0.251$\pm$0.090\\
170&5254 Ulysses (1986 VG1)&L4&9.2&0.368$\pm$0.025&0.384$\pm$0.052\\
171&5258 (1989 AU1)&L4&10.3&0.345$\pm$0.046&0.553$\pm$0.076\\
172&5259 Epeigeus (1989 BB1)&L4&10.4&0.503$\pm$0.025&0.291$\pm$0.084\\
173&5264 Telephus (1991 KC)&L4&9.5&0.386$\pm$0.037&0.431$\pm$0.085\\
174&5283 Pyrrhus (1989 BW)&L4&9.7&0.500$\pm$0.043&0.545$\pm$0.100\\
175&5284 Orsilocus (1989 CK2)&L4&10.1&0.447$\pm$0.027&0.379$\pm$0.085\\
176&5285 Krethon (1989 EO11)&L4&10.1&0.413$\pm$0.057&0.314$\pm$0.081\\
177&53436 (1999 VB154)&L4&11.4&0.361$\pm$0.071&0.559$\pm$0.141\\
178&5436 Eumelos (1990 DK)&L4&10.4&0.353$\pm$0.050&0.438$\pm$0.100\\
179&55571 (2002 CP82)&L4&12.0&0.321$\pm$0.064&0.548$\pm$0.135\\
180&5652 Amphimachus (1992 HS3)&L4&10.1&0.342$\pm$0.036&0.171$\pm$0.057\\
181&588 Achilles (A906 DN)&L4&8.3&0.450$\pm$0.041&0.341$\pm$0.083\\
182&60383 (2000 AR184)&L4&11.1&0.366$\pm$0.109&0.287$\pm$0.108\\
183&6090 (1989 DJ)&L4&9.4&0.434$\pm$0.033&0.054$\pm$0.080\\
184&624 Hektor (A907 CF)&L4&7.3&0.420$\pm$0.009&0.243$\pm$0.044\\
185&63273 (2001 DH4)&L4&11.5&0.285$\pm$0.072&0.348$\pm$0.186\\
186&63286 (2001 DZ68)&L4&12.1&0.529$\pm$0.183&0.081$\pm$0.160\\
187&65097 (2002 CC4)&L4&12.0&0.419$\pm$0.075&0.484$\pm$0.237\\
188&65257 (2002 FU36)&L4&11.5&0.402$\pm$0.098&0.577$\pm$0.268\\
189&6545 (1986 TR6)&L4&10.2&0.334$\pm$0.052&0.382$\pm$0.084\\
190&659 Nestor (A908 FE)&L4&8.7&0.263$\pm$0.030&-0.010$\pm$0.022\\
191&67065 (1999 XW261)&L4&11.9&0.331$\pm$0.061&1.936$\pm$0.765\\
192&7119 Hiera (1989 AV2)&L4&9.7&0.440$\pm$0.045&0.103$\pm$0.083\\
193&7152 Euneus (1973 SH1)&L4&10.3&0.365$\pm$0.034&0.285$\pm$0.067\\
194&7543 Prylis (1973 SY)&L4&10.6&0.329$\pm$0.033&0.297$\pm$0.105\\
195&7641 (1986 TT6)&L4&9.5&0.427$\pm$0.027&0.418$\pm$0.072\\
196&8125 Tyndareus (5493 T-2)&L4&10.8&0.355$\pm$0.035&0.418$\pm$0.088\\
197&8241 Agrius (1973 SE1)&L4&11.2&0.366$\pm$0.086&0.617$\pm$0.130\\
198&8317 Eurysaces (4523 P-L)&L4&11.1&0.234$\pm$0.063&0.088$\pm$0.115\\
199&90337 (2003 FQ97)&L4&11.6&0.192$\pm$0.058&0.459$\pm$0.119\\
200&911 Agamemnon (A919 FB)&L4&7.9&0.453$\pm$0.012&0.133$\pm$0.076\\
201&9431 (1996 PS1)&L4&10.6&0.415$\pm$0.028&0.545$\pm$0.105\\
202&9694 Lycomedes (6581 P-L)&L4&10.7&0.353$\pm$0.064&0.527$\pm$0.100\\
203&9712 Nauplius (1973 SO1)&L4&10.9&0.263$\pm$0.127&0.381$\pm$0.191\\
204&9713 Oceax (1973 SP1)&L4&11.3&0.408$\pm$0.191&0.465$\pm$0.171\\
205&9790 (1995 OK8)&L4&10.9&0.324$\pm$0.044&0.430$\pm$0.081\\
206&9799 (1996 RJ)&L4&9.7&0.424$\pm$0.041&0.326$\pm$0.056\\
207&9817 Thersander (6540 P-L)&L4&11.5&0.256$\pm$0.167&0.695$\pm$0.252\\
208&9818 Eurymachos (6591 P-L)&L4&11.0&0.244$\pm$0.061&0.292$\pm$0.097\\
209&9857 (1991 EN)&L4&10.3&0.394$\pm$0.057&0.240$\pm$0.096\\
210&11089 (1994 CS8)&L5&10.7&0.316$\pm$0.038&0.205$\pm$0.075\\
211&11487 (1988 RG10)&L5&11.3&0.349$\pm$0.076&0.426$\pm$0.102\\
212&11509 Thersilochos (1990 VL6)&L5&10.1&0.426$\pm$0.027&0.423$\pm$0.081\\
213&11552 Boucolion (1993 BD4)&L5&10.1&0.263$\pm$0.035&0.359$\pm$0.058\\
214&11554 Asios (1993 BZ12)&L5&10.5&0.324$\pm$0.016&0.336$\pm$0.050\\
215&11663 (1997 GO24)&L5&10.9&0.383$\pm$0.087&0.346$\pm$0.146\\
216&1172 Aneas (1930 UA)&L5&8.2&0.491$\pm$0.033&0.346$\pm$0.063\\
217&1173 Anchises (1930 UB)&L5&8.9&0.313$\pm$0.040&0.173$\pm$0.047\\
218&11887 Echemmon (1990 TV12)&L5&10.8&0.448$\pm$0.057&0.399$\pm$0.162\\
219&12052 Aretaon (1997 JB16)&L5&10.6&0.406$\pm$0.049&0.269$\pm$0.081\\
220&12126 (1999 RM11)&L5&10.1&0.332$\pm$0.050&0.640$\pm$0.076\\
221&128299 (2003 YL61)&L5&11.5&0.391$\pm$0.070&0.466$\pm$0.114\\
222&12929 (1999 TZ1)&L5&10.0&0.392$\pm$0.030&0.044$\pm$0.069\\
223&15502 (1999 NV27)&L5&10.0&0.325$\pm$0.027&0.273$\pm$0.053\\
224&15977 (1998 MA11)&L5&10.4&0.483$\pm$0.019&0.377$\pm$0.076\\
225&16070 (1999 RB101)&L5&9.7&0.371$\pm$0.020&0.285$\pm$0.047\\
226&16560 Daitor (1991 VZ5)&L5&10.7&0.261$\pm$0.042&0.160$\pm$0.048\\
227&16667 (1993 XM1)&L5&10.8&0.332$\pm$0.060&0.806$\pm$0.192\\
228&16956 (1998 MQ11)&L5&10.7&0.316$\pm$0.046&0.704$\pm$0.121\\
229&17171 (1999 NB38)&L5&10.5&0.437$\pm$0.034&0.623$\pm$0.082\\
230&17172 (1999 NZ41)&L5&10.8&0.427$\pm$0.042&0.292$\pm$0.054\\
231&17314 Aisakos (1024 T-1)&L5&10.9&0.335$\pm$0.024&0.370$\pm$0.060\\
232&17365 (1978 VF11)&L5&10.5&0.279$\pm$0.056&0.812$\pm$0.208\\
233&17419 (1988 RH13)&L5&11.3&0.258$\pm$0.041&0.418$\pm$0.130\\
234&17492 Hippasos (1991 XG1)&L5&10.1&0.358$\pm$0.022&0.509$\pm$0.109\\
235&18046 (1999 RN116)&L5&10.5&0.380$\pm$0.035&0.417$\pm$0.075\\
236&18054 (1999 SW7)&L5&10.8&0.317$\pm$0.044&0.373$\pm$0.065\\
237&18137 (2000 OU30)&L5&11.0&0.380$\pm$0.051&0.241$\pm$0.054\\
238&18278 Drymas (4035 T-3)&L5&11.4&0.345$\pm$0.059&0.895$\pm$0.272\\
239&18493 Demoleon (1996 HV9)&L5&10.7&0.335$\pm$0.053&0.251$\pm$0.125\\
240&1867 Deiphobus (1971 EA)&L5&8.3&0.426$\pm$0.022&0.257$\pm$0.050\\
241&1870 Glaukos (1971 FE)&L5&10.6&0.305$\pm$0.040&0.241$\pm$0.051\\
242&1871 Astyanax (1971 FF)&L5&11.2&0.290$\pm$0.057&0.429$\pm$0.091\\
243&1872 Helenos (1971 FG)&L5&10.8&0.242$\pm$0.038&0.123$\pm$0.087\\
244&1873 Agenor (1971 FH)&L5&10.1&0.379$\pm$0.029&0.484$\pm$0.090\\
245&19020 (2000 SC6)&L5&10.5&0.317$\pm$0.034&0.301$\pm$0.085\\
246&2207 Antenor (1977 QH1)&L5&8.9&0.305$\pm$0.021&0.137$\pm$0.028\\
247&22180 (2000 YZ)&L5&10.2&0.558$\pm$0.028&0.983$\pm$0.237\\
248&2223 Sarpedon (1977 TL3)&L5&9.1&0.492$\pm$0.023&0.608$\pm$0.082\\
249&2241 Alcathous (1979 WM)&L5&8.5&0.387$\pm$0.010&0.290$\pm$0.041\\
250&23549 Epicles (1994 ES6)&L5&11.7&0.196$\pm$0.072&0.655$\pm$0.167\\
251&2357 Phereclos (1981 AC)&L5&8.9&0.404$\pm$0.008&0.348$\pm$0.031\\
252&2363 Cebriones (1977 TJ3)&L5&8.9&0.306$\pm$0.035&-0.201$\pm$0.044\\
253&24446 (2000 PR25)&L5&11.0&0.334$\pm$0.047&0.423$\pm$0.152\\
254&24448 (2000 QE42)&L5&11.4&0.303$\pm$0.087&0.510$\pm$0.125\\
255&24451 (2000 QS104)&L5&10.3&0.335$\pm$0.029&0.926$\pm$0.157\\
256&24453 (2000 QG173)&L5&11.2&0.250$\pm$0.060&0.315$\pm$0.082\\
257&24454 (2000 QF198)&L5&11.4&0.232$\pm$0.102&0.297$\pm$0.106\\
258&24470 (2000 SJ310)&L5&10.9&0.425$\pm$0.049&0.348$\pm$0.156\\
259&24471 (2000 SH313)&L5&11.1&0.277$\pm$0.086&0.647$\pm$0.388\\
260&25883 (2000 RD88)&L5&11.3&0.393$\pm$0.063&0.144$\pm$0.097\\
261&2674 Pandarus (1982 BC3)&L5&9.1&0.473$\pm$0.033&0.419$\pm$0.059\\
262&2893 Peiroos (1975 QD)&L5&9.0&0.333$\pm$0.028&0.344$\pm$0.047\\
263&2895 Memnon (1981 AE1)&L5&10.1&0.246$\pm$0.031&0.121$\pm$0.048\\
264&29603 (1998 MO44)&L5&11.1&0.272$\pm$0.059&0.687$\pm$0.133\\
265&29976 (1999 NE9)&L5&11.0&0.351$\pm$0.043&0.310$\pm$0.056\\
266&30504 (2000 RS80)&L5&11.4&0.187$\pm$0.100&0.753$\pm$0.141\\
267&30506 (2000 RO85)&L5&11.0&0.335$\pm$0.049&0.288$\pm$0.097\\
268&30704 Phegeus (3250 T-3)&L5&11.3&0.298$\pm$0.068&0.421$\pm$0.129\\
269&30705 Idaios (3365 T-3)&L5&10.4&0.410$\pm$0.028&0.700$\pm$0.119\\
270&30942 Helicaon (1994 CX13)&L5&11.4&0.322$\pm$0.050&0.228$\pm$0.086\\
271&31342 (1998 MU31)&L5&10.5&0.390$\pm$0.038&0.359$\pm$0.075\\
272&31344 Agathon (1998 OM12)&L5&10.9&0.271$\pm$0.060&0.551$\pm$0.110\\
273&31819 (1999 RS150)&L5&11.4&0.345$\pm$0.078&0.521$\pm$0.116\\
274&32339 (2000 QA88)&L5&11.7&0.296$\pm$0.079&0.685$\pm$0.187\\
275&32397 (2000 QL214)&L5&11.5&0.315$\pm$0.105&0.148$\pm$0.139\\
276&3240 Laocoon (1978 VG6)&L5&10.2&0.372$\pm$0.041&0.567$\pm$0.078\\
277&32435 (2000 RZ96)&L5&11.2&0.331$\pm$0.078&0.714$\pm$0.194\\
278&32440 (2000 RC100)&L5&11.4&0.360$\pm$0.098&0.696$\pm$0.146\\
279&32464 (2000 SB132)&L5&11.6&0.322$\pm$0.094&0.316$\pm$0.189\\
280&32475 (2000 SD234)&L5&10.8&0.306$\pm$0.051&0.671$\pm$0.216\\
281&32482 (2000 ST354)&L5&11.0&0.285$\pm$0.061&0.070$\pm$0.072\\
282&32499 (2000 YS11)&L5&10.5&0.303$\pm$0.053&0.667$\pm$0.121\\
283&32501 (2000 YV135)&L5&11.0&0.444$\pm$0.049&0.091$\pm$0.118\\
284&32615 (2001 QU277)&L5&11.0&0.330$\pm$0.065&0.280$\pm$0.093\\
285&32811 Apisaon (1990 TP12)&L5&11.3&0.329$\pm$0.061&0.097$\pm$0.083\\
286&3317 Paris (1984 KF)&L5&8.4&0.402$\pm$0.010&0.241$\pm$0.035\\
287&3451 Mentor (1984 HA1)&L5&8.5&0.157$\pm$0.050&-0.006$\pm$0.059\\
288&34642 (2000 WN2)&L5&10.8&0.345$\pm$0.043&0.129$\pm$0.082\\
289&34746 (2001 QE91)&L5&9.9&0.217$\pm$0.039&0.125$\pm$0.067\\
290&3708 (1974 FV1)&L5&9.3&0.478$\pm$0.032&0.300$\pm$0.050\\
291&37519 Amphios (3040 T-3)&L5&11.1&0.402$\pm$0.050&0.229$\pm$0.117\\
292&42277 (2001 SQ51)&L5&12.1&0.378$\pm$0.057&0.413$\pm$0.090\\
293&4348 Poulydamas (1988 RU)&L5&9.6&0.234$\pm$0.022&0.134$\pm$0.046\\
294&4707 Khryses (1988 PY)&L5&10.6&0.357$\pm$0.046&0.647$\pm$0.101\\
295&4708 Polydoros (1988 RT)&L5&9.9&0.383$\pm$0.017&0.369$\pm$0.039\\
296&4709 Ennomos (1988 TU2)&L5&8.6&0.270$\pm$0.038&0.107$\pm$0.062\\
297&4715 (1989 TS1)&L5&9.8&0.348$\pm$0.050&0.465$\pm$0.064\\
298&4722 Agelaos (4271 T-3)&L5&10.0&0.418$\pm$0.035&0.529$\pm$0.050\\
299&4754 Panthoos (5010 T-3)&L5&10.0&0.392$\pm$0.018&0.436$\pm$0.037\\
300&4791 Iphidamas (1988 PB1)&L5&10.0&0.526$\pm$0.044&0.395$\pm$0.100\\
301&4792 Lykaon (1988 RK1)&L5&10.1&0.353$\pm$0.036&0.481$\pm$0.062\\
302&4805 Asteropaios (1990 VH7)&L5&10.1&0.340$\pm$0.022&0.235$\pm$0.046\\
303&4827 Dares (1988 QE)&L5&10.5&0.389$\pm$0.029&0.262$\pm$0.053\\
304&4828 Misenus (1988 RV)&L5&10.4&0.325$\pm$0.040&0.242$\pm$0.048\\
305&4829 Sergestus (1988 RM1)&L5&11.2&0.335$\pm$0.043&0.488$\pm$0.102\\
306&4832 Palinurus (1988 TU1)&L5&10.0&0.408$\pm$0.027&0.579$\pm$0.056\\
307&48438 (1989 WJ2)&L5&10.9&0.404$\pm$0.032&0.522$\pm$0.152\\
308&4867 Polites (1989 SZ)&L5&9.8&0.420$\pm$0.014&0.449$\pm$0.045\\
309&48764 (1997 JJ10)&L5&11.5&0.431$\pm$0.067&0.911$\pm$0.232\\
310&5119 (1988 RA1)&L5&10.3&0.395$\pm$0.044&0.464$\pm$0.070\\
311&5120 Bitias (1988 TZ1)&L5&10.3&0.344$\pm$0.030&0.500$\pm$0.107\\
312&5130 Ilioneus (1989 SC7)&L5&9.8&0.398$\pm$0.021&0.277$\pm$0.049\\
313&51364 (2000 SU333)&L5&11.6&0.279$\pm$0.049&0.421$\pm$0.122\\
314&51365 (2000 TA42)&L5&10.7&0.309$\pm$0.038&0.358$\pm$0.160\\
315&5144 Achates (1991 XX)&L5&9.0&0.414$\pm$0.014&0.437$\pm$0.055\\
316&51958 (2001 QJ256)&L5&11.5&-0.022$\pm$0.145&0.846$\pm$0.327\\
317&51962 (2001 QH267)&L5&11.5&0.125$\pm$0.103&0.977$\pm$0.194\\
318&5233 (1988 RL10)&L5&11.5&0.344$\pm$0.051&0.351$\pm$0.066\\
319&54656 (2000 SX362)&L5&10.7&0.400$\pm$0.045&0.284$\pm$0.088\\
320&5476 (1989 TO11)&L5&10.6&0.366$\pm$0.036&0.230$\pm$0.075\\
321&55060 (2001 QM73)&L5&11.1&0.342$\pm$0.047&0.745$\pm$0.123\\
322&5511 Cloanthus (1988 TH1)&L5&10.3&0.446$\pm$0.049&0.550$\pm$0.093\\
323&55419 (2001 TF19)&L5&11.2&0.297$\pm$0.041&0.478$\pm$0.096\\
324&5638 Deikoon (1988 TA3)&L5&10.5&0.386$\pm$0.034&0.265$\pm$0.054\\
325&5648 (1990 VU1)&L5&9.7&0.350$\pm$0.025&0.505$\pm$0.091\\
326&56968 (2000 SA92)&L5&11.6&0.346$\pm$0.105&0.939$\pm$0.239\\
327&58008 (2002 TW240)&L5&11.6&0.403$\pm$0.036&0.350$\pm$0.100\\
328&5907 (1989 TU5)&L5&11.1&0.334$\pm$0.044&0.522$\pm$0.079\\
329&6002 (1988 RO)&L5&10.5&0.423$\pm$0.038&0.387$\pm$0.088\\
330&617 Patroclus (A906 UL)&L5&8.2&0.287$\pm$0.017&0.299$\pm$0.045\\
331&68444 (2001 RH142)&L5&11.5&0.244$\pm$0.085&0.734$\pm$0.158\\
332&6997 Laomedon (3104 T-3)&L5&10.6&0.336$\pm$0.055&0.143$\pm$0.087\\
333&7352 (1994 CO)&L5&10.0&0.308$\pm$0.034&0.152$\pm$0.043\\
334&76857 (2000 WE132)&L5&11.0&0.338$\pm$0.061&0.673$\pm$0.227\\
335&7815 Dolon (1987 QN)&L5&10.2&0.378$\pm$0.024&0.315$\pm$0.059\\
336&82055 (2000 TY40)&L5&11.7&0.304$\pm$0.105&0.363$\pm$0.123\\
337&884 Priamus (A917 SU)&L5&8.7&0.438$\pm$0.017&0.376$\pm$0.044\\
338&9023 Mnesthus (1988 RG1)&L5&10.2&0.375$\pm$0.032&0.233$\pm$0.048\\
339&9030 (1989 UX5)&L5&11.1&0.384$\pm$0.046&0.608$\pm$0.088\\
340&9142 Rhesus (5191 T-3)&L5&10.6&0.388$\pm$0.049&0.491$\pm$0.098\\
341&9430 Erichthonios (1996 HU10)&L5&11.3&0.352$\pm$0.046&0.886$\pm$0.148\\
342&99943 (2005 AS2)&L5&11.6&0.276$\pm$0.091&0.134$\pm$0.083\\

	\enddata
	\tablenotetext{a}{H magnitude was obtained from \url{https://ssd.jpl.nasa.gov/horizons.cgi}}
	
\end{deluxetable*}

\begin{deluxetable*}{|l|c|rr|r|rr|r|}
	\tablecaption{ATLAS objects with derived rotation periods. The listed false-alarm probability refers only to the confidence of a signal in the data at that frequency, this can not distinguish an alias from the true. period. \label{table_rot_data}}
	\tablehead{
		\colhead{Object Name} & \colhead{Trojan Group} & \colhead{Rot. Period } & \colhead{False Alarm Prob. } & \colhead{LCDB Period} & \colhead{Alias} & \colhead{Alias Period} &  \colhead{Amplitude}\\
		\colhead{} & \colhead{} & \colhead{(hours)} & \colhead{(prob.)} & \colhead{(hours)} & \colhead{(days)} & \colhead{(hours)} &  \colhead{(mag)}		
	}
	\startdata
	13183 (1996 TW)&L4&12.397$\pm$0.009&5.9e-17&-&-&-&0.286$\pm$0.044\\
15527 (1999 YY2)&L4&5.411$\pm$0.002&2.4e-22&6.990&-2.0&6.986&0.304$\pm$0.072\\
2920 Automedon (1981 JR)&L4&10.214$\pm$0.006&9.3e-16&10.212&-&-&0.232$\pm$0.036\\
3063 Makhaon (1983 PV)&L4&5.759$\pm$0.002&3.2e-11&-&-&-&0.080$\pm$0.019\\
3391 Sinon (1977 DD3)&L4&8.135$\pm$0.004&2.8e-13&8.135&-&-&0.676$\pm$0.107\\
3596 Meriones (1985 VO)&L4&12.835$\pm$0.010&2.2e-31&-&-&-&0.207$\pm$0.031\\
3793 Leonteus (1985 TE3)&L4&5.031$\pm$0.002&1.1e-16&5.621&-1.0&5.621&0.202$\pm$0.055\\
4063 Euforbo (1989 CG2)&L4&8.846$\pm$0.004&1.1e-13&8.846&-&-&0.207$\pm$0.031\\
4068 Menestheus (1973 SW)&L4&11.033$\pm$0.006&1.2e-28&-&-&-&0.238$\pm$0.045\\
4489 (1988 AK)&L4&9.969$\pm$0.009&8.7e-16&12.582&-1.0&12.582&0.155$\pm$0.039\\
5209 (1989 CW1)&L4&9.339$\pm$0.006&5.1e-18&-&-&-&0.327$\pm$0.048\\
5244 Amphilochos (1973 SQ1)&L4&9.786$\pm$0.006&3e-27&9.766&-&-&0.443$\pm$0.050\\
5264 Telephus (1991 KC)&L4&9.520$\pm$0.006&1.7e-33&9.525&-&-&0.527$\pm$0.104\\
5285 Krethon (1989 EO11)&L4&12.024$\pm$0.013&2.9e-26&-&-&-&0.416$\pm$0.031\\
7641 (1986 TT6)&L4&27.795$\pm$0.044&1.8e-55&27.770&-&-&0.377$\pm$0.058\\
9431 (1996 PS1)&L4&19.862$\pm$0.023&8.3e-22&-&-&-&0.437$\pm$0.082\\
11089 (1994 CS8)&L5&8.405$\pm$0.006&4.2e-24&7.720&0.5&7.729&0.431$\pm$0.063\\
1172 Aneas (1930 UA)&L5&8.703$\pm$0.007&4.8e-36&8.705&-&-&0.159$\pm$0.028\\
1173 Anchises (1930 UB)&L5&11.609$\pm$0.007&3e-26&11.595&-&-&0.513$\pm$0.073\\
16560 Daitor (1991 VZ5)&L5&13.800$\pm$0.008&3.5e-29&-&-&-&0.407$\pm$0.056\\
17365 (1978 VF11)&L5&12.672$\pm$0.019&1.6e-22&12.672&-&-&0.781$\pm$0.094\\
17492 Hippasos (1991 XG1)&L5&12.950$\pm$0.007&8.8e-23&-&-&-&0.404$\pm$0.082\\
1867 Deiphobus (1971 EA)&L5&59.239$\pm$0.353&1.5e-41&-&-&-&0.261$\pm$0.035\\
1872 Helenos (1971 FG)&L5&5.810$\pm$0.002&8.4e-16&-&-&-&0.593$\pm$0.089\\
2207 Antenor (1977 QH1)&L5&4.776$\pm$0.001&4.5e-12&-&-&-&0.227$\pm$0.036\\
2674 Pandarus (1982 BC3)&L5&8.478$\pm$0.005&9.1e-48&8.480&-&-&0.537$\pm$0.056\\
2893 Peiroos (1975 QD)&L5&8.949$\pm$0.005&1e-38&8.945&-&-&0.360$\pm$0.058\\
2895 Memnon (1981 AE1)&L5&7.520$\pm$0.002&2.8e-50&7.516&-&-&0.497$\pm$0.066\\
30705 Idaios (3365 T-3)&L5&13.512$\pm$0.009&1.3e-17&15.736&-0.5&15.725&0.386$\pm$0.068\\
3240 Laocoon (1978 VG6)&L5&11.313$\pm$0.008&2.2e-35&-&-&-&0.466$\pm$0.071\\
32615 (2001 QU277)&L5&6.716$\pm$0.004&6.7e-16&6.712&-&-&0.339$\pm$0.070\\
3317 Paris (1984 KF)&L5&7.081$\pm$0.002&2.3e-26&7.081&-&-&0.088$\pm$0.023\\
3451 Mentor (1984 HA1)&L5&7.697$\pm$0.002&1.9e-96&7.702&-&-&0.670$\pm$0.033\\
4348 Poulydamas (1988 RU)&L5&8.219$\pm$0.003&8.8e-35&9.908&-1.0&9.917&0.310$\pm$0.054\\
4709 Ennomos (1988 TU2)&L5&12.270$\pm$0.010&2.9e-17&12.275&-&-&0.456$\pm$0.038\\
4715 (1989 TS1)&L5&8.814$\pm$0.003&5.2e-80&8.813&-&-&0.428$\pm$0.043\\
4827 Dares (1988 QE)&L5&18.958$\pm$0.018&4.4e-19&18.995&-&-&0.261$\pm$0.069\\
4828 Misenus (1988 RV)&L5&12.858$\pm$0.014&4.8e-59&12.873&-&-&0.374$\pm$0.053\\
5144 Achates (1991 XX)&L5&6.799$\pm$0.004&2.1e-20&5.958&1.0&5.956&0.169$\pm$0.044\\
884 Priamus (A917 SU)&L5&6.861$\pm$0.003&1.1e-21&6.861&-&-&0.239$\pm$0.032\\
9030 (1989 UX5)&L5&6.307$\pm$0.003&3.2e-12&-&-&-&0.454$\pm$0.081\\

	\enddata
	
\end{deluxetable*}



\end{document}